\documentclass[ALICE,manyauthors]{cernphprep}
\usepackage[comma,square,numbers,sort&compress]{natbib}
\usepackage{hyperref}
\usepackage{lineno}
\usepackage{xspace}
\usepackage{booktabs} 
\usepackage{color}
\usepackage[T1]{fontenc}
\usepackage{orcidlink}

\begin{document}
%

\newcommand{\pp}           {pp\xspace}
\newcommand{\ppbar}        {\mbox{$\mathrm {p\overline{p}}$}\xspace}
\newcommand{\XeXe}         {\mbox{Xe--Xe}\xspace}
\newcommand{\PbPb}         {\mbox{Pb--Pb}\xspace}
\newcommand{\pA}           {\mbox{pA}\xspace}
\newcommand{\pPb}          {\mbox{p--Pb}\xspace}
\newcommand{\AuAu}         {\mbox{Au--Au}\xspace}
\newcommand{\dAu}          {\mbox{d--Au}\xspace}

\newcommand{\s}            {\ensuremath{\sqrt{s}}\xspace}
\newcommand{\snn}          {\ensuremath{\sqrt{s_{\mathrm{NN}}}}\xspace}
\newcommand{\pt}           {\ensuremath{p_{\rm T}}\xspace}
\newcommand{\pte}          {\ensuremath{p_{\rm T,e}}\xspace}   
\newcommand{\ptee}         {\ensuremath{p_{\rm T,ee}}\xspace}  
\newcommand{\mee}          {\ensuremath{m_{\rm ee}}\xspace}     
\newcommand{\mt}           {\ensuremath{m_{\rm T}}\xspace}     
\newcommand{\meanpt}       {$\langle p_{\mathrm{T}}\rangle$\xspace}
\newcommand{\ycms}         {\ensuremath{y_{\rm CMS}}\xspace}
\newcommand{\ylab}         {\ensuremath{y_{\rm lab}}\xspace}
\newcommand{\etarange}[1]  {\mbox{$\left | \eta \right |~<~#1$}}
\newcommand{\yrange}[1]    {\mbox{$\left | y \right |~<~#1$}}
\newcommand{\dndy}         {\ensuremath{\mathrm{d}N_\mathrm{ch}/\mathrm{d}y}\xspace}
\newcommand{\dndeta}       {\ensuremath{\mathrm{d}N_\mathrm{ch}/\mathrm{d}\eta}\xspace}
\newcommand{\avdndeta}     {\ensuremath{\langle\dndeta\rangle}\xspace}
\newcommand{\dNdy}         {\ensuremath{\mathrm{d}N_\mathrm{ch}/\mathrm{d}y}\xspace}
\newcommand{\Npart}        {\ensuremath{N_\mathrm{part}}\xspace}
\newcommand{\Ncoll}        {\ensuremath{N_\mathrm{coll}}\xspace}
\newcommand{\dEdx}         {\ensuremath{\textrm{d}E/\textrm{d}x}\xspace}
\newcommand{\RpPb}         {\ensuremath{R_{\rm pPb}}\xspace}

\newcommand{\nineH}        {$\sqrt{s}~=~0.9$~Te\kern-.1emV\xspace}
\newcommand{\seven}        {$\sqrt{s}~=~7$~Te\kern-.1emV\xspace}
\newcommand{\twoH}         {$\sqrt{s}~=~0.2$~Te\kern-.1emV\xspace}
\newcommand{\twosevensix}  {$\sqrt{s}~=~2.76$~Te\kern-.1emV\xspace}
\newcommand{\five}         {$\sqrt{s}~=~5.02$~Te\kern-.1emV\xspace}
\newcommand{\twosevensixnn}{$\sqrt{s_{\mathrm{NN}}}~=~2.76$~Te\kern-.1emV\xspace}
\newcommand{\fivenn}       {$\sqrt{s_{\mathrm{NN}}}~=~5.02$~Te\kern-.1emV\xspace}
\newcommand{\LT}           {L{\'e}vy-Tsallis\xspace}
\newcommand{\GeVc}         {Ge\kern-.1emV/$c$\xspace}
\newcommand{\MeVc}         {Me\kern-.1emV/$c$\xspace}
\newcommand{\TeV}          {Te\kern-.1emV\xspace}
\newcommand{\GeV}          {Ge\kern-.1emV\xspace}     
\newcommand{\MeV}          {Me\kern-.1emV\xspace}
\newcommand{\GeVmass}      {Ge\kern-.1emV/$c^2$\xspace}
\newcommand{\MeVmass}      {Me\kern-.1emV/$c^2$\xspace}
\newcommand{\lumi}         {\ensuremath{\mathcal{L}}\xspace}
\newcommand\tenpow[1]{\ensuremath{{\times}10^{#1}}} 
\newcommand{\pbarn}        {pb$^{-1}$\xspace} 
\newcommand{\nbarn}        {nb$^{-1}$\xspace} 

\newcommand{\ITS}          {\rm{ITS}\xspace}
\newcommand{\TOF}          {\rm{TOF}\xspace}
\newcommand{\ZDC}          {\rm{ZDC}\xspace}
\newcommand{\ZDCs}         {\rm{ZDCs}\xspace}
\newcommand{\ZNA}          {\rm{ZNA}\xspace}
\newcommand{\ZNC}          {\rm{ZNC}\xspace}
\newcommand{\SPD}          {\rm{SPD}\xspace}
\newcommand{\SDD}          {\rm{SDD}\xspace}
\newcommand{\SSD}          {\rm{SSD}\xspace}
\newcommand{\TPC}          {\rm{TPC}\xspace}
\newcommand{\TRD}          {\rm{TRD}\xspace}
\newcommand{\VZERO}        {\rm{V0}\xspace}
\newcommand{\VZEROA}       {\rm{V0A}\xspace}
\newcommand{\VZEROC}       {\rm{V0C}\xspace}
\newcommand{\Vdecay} 	   {\ensuremath{V^{0}}\xspace}

\newcommand{\ee}           {\ensuremath{e^{+}e^{-}}} 
\newcommand{\pip}          {\ensuremath{\pi^{+}}\xspace}
\newcommand{\pim}          {\ensuremath{\pi^{-}}\xspace}
\newcommand{\piz}          {\ensuremath{\pi^{0}}\xspace}  
\newcommand{\etap}         {\ensuremath{\eta^{\prime}}\xspace}    
\newcommand{\rhom}         {\ensuremath{\rho}\xspace}     
\newcommand{\om}           {\ensuremath{\omega}\xspace}   
\newcommand{\phim}         {\ensuremath{\phi}\xspace}     
\newcommand{\jpsi}         {\ensuremath{J/\psi}\xspace}   
\newcommand{\etapiz}       {\ensuremath{\eta/\pi^{0}}\xspace}    

\newcommand{\kap}          {\ensuremath{\rm{K}^{+}}\xspace}
\newcommand{\kam}          {\ensuremath{\rm{K}^{-}}\xspace}
\newcommand{\pbar}         {\ensuremath{\rm\overline{p}}\xspace}
\newcommand{\kzero}        {\ensuremath{{\rm K}^{0}_{\rm{S}}}\xspace}
\newcommand{\lmb}          {\ensuremath{\Lambda}\xspace}
\newcommand{\almb}         {\ensuremath{\overline{\Lambda}}\xspace}
\newcommand{\Om}           {\ensuremath{\Omega^-}\xspace}
\newcommand{\Mo}           {\ensuremath{\overline{\Omega}^+}\xspace}
\newcommand{\X}            {\ensuremath{\Xi^-}\xspace}
\newcommand{\Ix}           {\ensuremath{\overline{\Xi}^+}\xspace}
\newcommand{\Xis}          {\ensuremath{\Xi^{\pm}}\xspace}
\newcommand{\Oms}          {\ensuremath{\Omega^{\pm}}\xspace}
\newcommand{\degree}       {\ensuremath{^{\rm o}}\xspace}

\newcommand{\gdir}         {\gamma_{\rm{dir}}} 
\newcommand{\gincl}        {\gamma_{\rm{inc}}} 
\newcommand{\gdecay}       {\gamma_{\rm{dec}}} 
\newcommand{\geantthree}       {\ensuremath{\textsc{Geant3}}\xspace}

\begin{titlepage}
\PHyear{2024}       
\PHnumber{295}      
\PHdate{11 November}  

\title{Direct-photon production in inelastic and high-multiplicity \ proton--proton collisions at $\sqrt{s} =$ 13 TeV}
\ShortTitle{Direct photons in pp 13 TeV}   

\Collaboration{ALICE Collaboration\thanks{See Appendix~\ref{app:collab} for the list of collaboration members}}
\ShortAuthor{ALICE Collaboration} 

\begin{abstract}
In this letter, we present the first measurement of direct photons at the transverse momentum of $ 1 < \pt < 6$ \GeVc at midrapidity $|\eta| < 0.8$ in inelastic and high-multiplicity proton--proton collisions at a centre-of-mass energy of $\sqrt{s} =$ 13 TeV. The fraction of virtual direct photons in the inclusive virtual photon spectrum is obtained from a fit to the dielectron invariant mass spectrum. In the limit of zero invariant mass, this fraction is equal to the relative contribution of real direct photons in the inclusive real photon spectrum. Contributions from decays of light-flavour neutral mesons are estimated using independent measurements in proton--proton collisions at the same energy and the same event class. For the first time at the LHC energies, a direct-photon signal is observed at low \pt in both inelastic and high-multiplicity event classes, with a significance of 3.2$\sigma$ and 1.9$\sigma$ in terms of standard deviations, correspondingly. The yield of direct photons in inelastic pp collisions is compared to perturbative QCD calculations. The integrated photon yield is studied as a function of charged-particle multiplicity and is compared to the results from other experiments and theoretical calculations. The results show a significant increase of direct-photon yield with charged-particle multiplicity.
\end{abstract}

\end{titlepage}

\setcounter{page}{2} 


\section{Introduction}

Extensive studies of ultra-relativistic heavy-ion collisions in the last decades have established the formation of a quark--gluon plasma (QGP) in these collisions, a deconfined state of matter in which quarks and gluons are not bound inside nucleons, see e.g.~Refs.~\cite{ALICE:2022wpn,Braun-Munzinger:2015hba,Busza:2018rrf} for recent reviews.  
Experimental evidence for the creation of a strongly-interacting QGP has been found in many observations, including collective phenomena described by relativistic hydrodynamics~\cite{Voloshin:2008dg,kolb2004hydrodynamic}, energy loss effects of colour-charged probes~\cite{Wiedemann:1200905}, sequential suppression and (re)generation of quarkonium states~\cite{CMS:2012gvv}, and electromagnetic radiation consistent with high initial temperatures of several hundred MeV~\cite{PHENIX:2008uif,ALICE:2015xmh,ALICE:2023jef}. The latter signature, thermal photons (i.e.~photons produced by a thermalised medium), plays an important role in the studies of the hot system created in heavy-ion collisions~\cite{Ruuskanen:1991au,Geurts:2022xmk,David:2019wpt}. While the momentum distribution of direct photons (i.e.~photons not originating from hadron decays) is defined by the kinematics of interacting partons, the momentum distribution of thermally produced direct photons is related to the medium temperature, with a black-body-type radiation spectrum following an exponential law $e^{-E/T}$, and blue-shifted by radial flow effects from the rapidly expanding medium~\cite{vanHees:2011vb,dEnterria:2005jvq,Shen:2013vja}. In general, direct photons can be created both in the form of real (massless) photons and in the form of virtual photons with non-zero mass~\cite{Dalitz:1951aj}, which convert into a lepton pair (dimuon $\mu^{+}\mu^{-}$ or dielectron e$^{+}$e$^{-}$). An important class of direct photons are so-called prompt photons produced in initial hard scattering processes between the partons such as quark--antiquark annihilation and quark--gluon Compton scattering. At the transverse momentum of \mbox{\pt $>$ 3 \GeVc} this contribution can be well described by perturbative quantum chromodynamics (pQCD) calculations~\cite{Gordon:1993qc}. 

Smaller systems such as proton--proton (pp) and proton--nucleus collisions in which a large number of charged particles are produced have recently attracted the interest of the heavy-ion community~\cite{Nagle:2018nvi,Dusling:2015gta,Shuryak:2013ke}. These events exhibit several features which were previously attributed to heavy-ion collisions, including collective effects such as long-range angular correlations and flow~\cite{Grosse-Oetringhaus:2024bwr,Schlichting:2016sqo,CMS:2010ifv,ALICE:2019zfl,ATLAS:2015hzw,CMS:2016fnw} or enhanced strangeness production~\cite{ALICE:2016fzo}. Moreover, also heavy-quark production is found to scale faster than linearly with the charged-particle multiplicity~\cite{ALICE:2015ikl}, but no evidence for energy loss effects of high-momentum probes has been observed~\cite{ATLAS:2022iyq,ALICE:2017svf,ALICE:2023plt}. These observations have sparked active discussions about whether the physics relevant to heavy-ion collisions is also at work in smaller collision systems, what conditions are necessary for nuclear matter to form (a droplet of) QGP, or whether it is possible to explain the observed phenomena through, for example, multiple parton interactions arising from colour reconnection effects~\cite{OrtizVelasquez:2013ofg,Bierlich:2014xba}. One can argue that the combination of a small volume created in pp collisions and a high multiplicity of produced charged particles should imply a very high energy density system. Studies of small systems should, therefore, provide crucial information needed to establish the transition from ``micro'' to ``macro'' dynamical regimes, which are often treated by quite different theoretical means~\cite{Nagle:2018nvi,Dusling:2015gta}.

If a thermalised system is produced in such collisions, a signal of thermal radiation in the form of (virtual) direct photons should be present and contribute to the photon spectrum, especially in the low transverse momentum region of \pt $<$ 3 \GeVc~\cite{Shen:2016zpp}. In this momentum range, pQCD calculations are challenging, so measurements of the direct photon yield in pp collisions have often been used as a reference for thermal radiation from the QGP created in heavy-ion collisions~\cite{PHENIX:2008uif,ALICE:2015xmh,STAR:2016use,PHENIX:2015vek}. However, a possible presence of thermal radiation in pp collisions could complicate the correct interpretation of these results obtained in heavy-ion collisions. At higher transverse momentum of \pt $\gtrsim$ 3 \GeVc, the measurement of direct photons in inelastic pp collisions also provides a valuable test for pQCD calculations. Compared to the RHIC energies, at the LHC energies the contribution from higher-order processes plays a much more important role than that of leading order~\cite{Gordon:1993qc}, so the experimental input on direct-photon production is relevant for theoretical models.

So far, the studies of real photons in proton--proton and proton--nucleus collisions at the LHC energies did not reveal significant contributions from direct photons at low transverse momentum~\cite{ALICE:2018mjj,ALICE:2019rtd}. Such studies can be based on the measurements of real photons in electromagnetic calorimeters, or they can be performed by tracking external conversions of real photons to e$^{+}$e$^{-}$ pairs in the detector material. However, these measurements suffer from a large background due to $\pi^{0} \rightarrow \gamma \gamma$ decays 
with a branching ratio of 98.8\%~\cite{ParticleDataGroup:2022pth}. This background can be largely suppressed by measuring the contribution from virtual direct photons~\cite{PHENIX:2009gyd,PHENIX:2018che,PHENIX:2012krx}, which provide an additional observable compared to real photon measurement --- the invariant mass of the created dielectron pair \mee. The relation between the real photon yield and the associated dielectron pair production is described by the Kroll-Wada equation~\cite{Kroll:1955zu,Landsberg:1985gaz}, which has a $\sim 1/m_{\rm ee}$ dependence of the corresponding dielectron spectrum for $p_{\rm T,ee} \gg m_{\rm ee}$. The measurement of virtual direct photons allows one to select the mass range $m_{\rm ee} \gtrsim m_{\pi^{0}} = 135$ \MeVmass, which drastically improves the signal-to-background ratio compared to measurements of real direct photons. The drawbacks of this method are the rapidly decreasing cross section as a function of invariant mass ($\sim 1/m_{\rm ee}$) and the small internal conversion probability of virtual photons ($\sim \alpha_{EM} \approx 1/137$).

In this letter we report the first measurement of direct photons in inelastic (INEL) and high-multiplicity (HM) pp collisions at $\sqrt{s} =$ 13 TeV at low transverse momentum of \mbox{$1 < \pt < 6$ \GeVc} at midrapitidy $|\eta| < 0.8$ using the virtual photon method. The analysis technique and the methods to extract the direct-photon signal follow the previous publication~\cite{ALICE:2018gev}, however the present analysis utilises an approximately four times larger statistics of pp collisions and the recent measurements of $\pi^{0}$ and $\eta$ yields in the same event classes~\cite{ALICE:2024vgi}. The paper is organised as follows: the ALICE apparatus and the data samples are described in Section~\ref{sec:alice}, the data analysis and the ``cocktail'' of known hadronic sources are discussed in Section~\ref{sec:data}, and the results are presented in Section~\ref{sec:results}.

\section{The ALICE detector and data samples}
\label{sec:alice}

A detailed description of the ALICE apparatus and its performance can be found in Refs.~\cite{ALICE:2004fvi,ALICE:2005vhb,ALICE:2008ngc,ALICE:2014sbx}. Below, we briefly describe the detectors used in this analysis to reconstruct and identify dielectrons.

The reconstruction of charged-particle tracks in the ALICE central barrel relies on the information from the Inner Tracking System (ITS)~\cite{ALICE:2010tia} and the Time Projection Chamber (TPC)~\cite{Alme:2010ke}. Both of these detectors are located inside a solenoid, providing a uniform magnetic field of 0.5 T parallel to the beam axis. The ITS comprises 6 cylindrical layers of silicon detectors located at radial distances from the beam axis between 3.9 cm and 43 cm. The ITS is surrounded by the TPC, the main tracking device in the ALICE central barrel. It is a 5 m long cylindrical gaseous detector extending from 85 cm to 247 cm in the radial direction. It provides up to 159 spacial points per track for charged-particle reconstruction and particle identification (PID). The latter relies on the measurement of the specific energy loss d$E$/d$x$ in the gas volume, which was filled with an Ar/CO$_{2}$ gas mixture (with abundances of 88/12) in 2016 and 2018 and with Ne/CO$_{2}$/N$_{2}$ (90/10/5) in 2017. The PID is complemented by the Time-of-Flight (TOF)~\cite{ALICE:2000xcm} system placed at a radial distance of 3.7 m from the beam axis. It provides the arrival time of particles relative to the event collision time. The latter can be measured either by the TOF detector itself or by the T0 detector comprising two arrays of Cherenkov counters at forward rapidities~\cite{ALICE:2016ovj}.

Collision events are triggered by the V0 detector, a set of two plastic scintillator arrays placed on both sides of the interaction point at forward rapidities. Together with the two innermost layers of the ITS detector, the V0 is also used to reject background events such as beam--gas interactions, collisions with de-bunched protons or with mechanical structures of the beam line. The analysis employs the full statistics of pp collisions at $\sqrt{s} = 13$ TeV recorded during the Run 2 of the LHC operation between 2015 and 2018. Simultaneous signals in both V0 arrays synchronised with the beam crossing time are used as a minimum-bias (MB) trigger, that samples $72\%$ of the total inelastic (INEL) cross section~\cite{ALICE-PUBLIC-2021-005, TOTEM:2017asr}. High charged-particle multiplicity events are triggered by the additional requirement that the total V0 signal amplitude exceeds a certain threshold. At the analysis level, $0.1\%$ of MB events (i.e.~$0.072\%$ of all INEL events) with the highest V0 multiplicity are selected to define the high-multiplicity class. The corresponding charged-particle multiplicity density at midrapidity amounts to $\dndeta = 5.31 \pm 0.18$ for INEL and to $\dndeta = 31.34 \pm 0.52$ for HM events~\cite{ALICE:2015qqj,ALICE:2020swj}. The primary interaction vertex is reconstructed with the track segments in the two innermost layers of the ITS detector, and pileup events (with multiple reconstructed vertices) are removed from the analysis. The vertex position along the beam axis is required to be within $\pm 10$ cm of the nominal interaction point to ensure a uniform detector coverage in the $|\eta| < 0.8$ range. In total 1.73 \tenpow{9} minimum-bias and \mbox{3.38 \tenpow{8}} high-multiplicity \pp events are selected for the analysis, corresponding to integrated luminosities of \mbox{\ensuremath{\mathcal{L}^{\rm {MB}}_{\rm{int}}}\xspace = $29.8 \pm 0.6$ \nbarn} and \ensuremath{\mathcal{L}^{\rm {HM}}_{\rm{int}}}\xspace = $5.8 \pm 0.1$ \pbarn, respectively. 
The luminosity determination is based on the visible cross section observed by the V0 trigger measured in a van der Meer scan~\cite{ALICE-PUBLIC-2021-005}.

\section{Data analysis and cocktail of known hadronic sources}
\label{sec:data}

Electron and positron candidates are selected in the transverse momentum range $\pte > $ 0.2 \GeVc and pseudorapidity range $|\eta_{\rm e}| < $ 0.8. Track quality and PID criteria are the same as those used in the previous analysis~\cite{ALICE:2018gev}. The electron and positron identification is based on the complementary information from the TPC and the TOF detectors, where the latter suppresses kaon and proton contamination in the momentum ranges in which the TPC \dEdx signals of these hadrons overlap with those of electrons. All electron and positron candidates are paired to form combinations with opposite ($N_{+-}$) and same-sign charges ($N_{\pm \pm}$). As in the previous publication~\cite{ALICE:2018gev}, the combinatorial background $B$ is estimated via the geometric mean of same-sign pairs within the same event $\sqrt{N_{++}N_{--}}$ and is corrected for the slightly different detector acceptance of opposite and same-sign pairs. The corresponding correction factor $R$ is estimated using uncorrelated opposite ($M_{+-}$) and same-sign ($M_{\pm \pm}$) pairs as $R = M_{+-}/(2\sqrt{M_{++}M_{--}})$. The different tracks in each of these pairs are taken from different (``mixed'') events to make sure that the $R$ factor reflects only the detector acceptance. The raw dielectron signal is then obtained as $S = N_{+-} - 2R\sqrt{N_{++}N_{--}}$. As in the previous analysis~\cite{ALICE:2018gev}, the resulting signal-to-background ratio has a minimum of around \mbox{$m_{ee} \sim 0.5$ \GeVmass} and ranges between 0.5 and 0.1 (between 0.2 and 0.03) for INEL (HM) events in the mass range of $0.14 < \mee < 0.35$ \GeVmass relevant for the direct-photon analysis. Dielectrons from real photon conversions in the detector material are removed from the analysis sample in the mass range below 100 \MeVmass by using their distinct orientation with respect to the magnetic field. The data are corrected for reconstruction efficiency using detailed Monte Carlo (MC) simulations of the ALICE detector performance. For this purpose, pp events are generated using the Monash 2013 tune of \textsc{Pythia} 8~\cite{Sjostrand:2014zea} for light-flavour (LF) decays and the Perugia 2011 tune of \textsc{Pythia}~6.4~\cite{Sjostrand:2006za} for heavy-flavour (HF) decays. The \jpsi decays are simulated using PHOTOS~\cite{Barberio:1993qi}, including the radiative decay \jpsi $\rightarrow \gamma~\rm{e}^{+}\rm{e}^{-}$. Generated particles are then propagated through the detector material using the \geantthree toolkit~\cite{Brun:1994aa}. The reconstruction efficiency is studied as a function of invariant mass and pair transverse momentum separately for different sources of dielectron pairs. The total signal efficiency amounts to $\sim 25\%$ in the mass range of $0.14 < \mee < 0.35$ \GeVmass used to extract the direct-photon fraction. This efficiency is obtained by weighting the efficiency of each source based on the hadronic cocktail simulation as described in the following.

Systematic uncertainties of the data are evaluated as described in Refs.~\cite{ALICE:2018gev,ALICE:2020umb} and are summarised in Table~\ref{tab:mee_syst} of the Appendix~\ref{sec:supplemental}. This includes the estimation of the corresponding uncertainty related to tracking such as selection criteria in the ITS and TPC detectors, ITS--TPC track matching, the requirement of a hit in the first ITS layer, and the requirement of no shared clusters in the ITS. The relative disagreements found between data and MC simulations are assigned as systematic uncertainty at the single-track level, which is doubled in the case of uncertainty for pairs. The MC simulations were also found to reproduce the details of the PID selection well, with a negligible resulting hadron contamination on the dielectron signal. At low invariant mass (\mbox{$m_{\rm ee} < 0.1~{\rm GeV}/c^{2}$}), the uncertainty on the conversion rejection was estimated by varying the criteria used to reject photon conversions. The uncertainty of the background estimation was estimated by repeating the event mixing in different event classes, defined by the position of the reconstructed primary vertex and by the charged-particle multiplicity, and by re-evaluating the correction factor $R$ in the raw signal calculation. A possible multiplicity dependence of the reconstruction and PID efficiency is covered by an uncertainty of up to 6\%~\cite{ALICE:2013wgn, ALICE:2020nkc}. The resulting total systematic uncertainty of the efficiency-corrected dielectron yield is 6.5--8.1$\%$ in INEL and to 6.8--12$\%$ in the HM analysis, which is assumed to be fully correlated as a function of dielectron mass \mee. The main improvement over the previous analysis~\cite{ALICE:2018gev} (with $\sim$14--15$\%$ systematic uncertainty) is due to better tracking from the more recent reconstruction of Run 2 data.

The measured dielectron yield is compared to the total sum (``cocktail'') of known hadronic sources, i.e.~all hadronic decays resulting in dielectron pairs in the final state, using a fast MC simulation as described in more detail in Refs.~\cite{ALICE:2018fvj, ALICE:2018gev}. The comparison is made within the ALICE central barrel acceptance ($|\eta_{\rm e}| < 0.8$ and $p_{\rm{T,e}} > 0.2$ \GeVc) and includes decays from $\pi^{0}$, $\eta$, $\eta'$, $\rho$, $\omega$ and $\phi$ mesons in the LF sector and \jpsi and correlated semi-leptonic decays in the HF sector. The detector resolution effects, including momentum and angular resolution as well as bremsstrahlung obtained from full MC simulations, are applied to the generated dielectrons as in Ref.~\cite{ALICE:2018gev}. Among these sources, first and foremost $\pi^{0}$ and $\eta$ mesons have a major contribution to the dielectron spectrum in the invariant mass range $\mee < 0.35$ \GeVmass used in this analysis to extract the fraction of (virtual) direct photons. Recent measurements of these mesons in INEL and HM pp collisions at $\sqrt{s} = 13~\rm{TeV}$~\cite{ALICE:2024vgi} are parameterised with a modified Hagedorn function~\cite{Hagedorn:1965st} in order to be used as an input for the fast MC simulation. These precise measurements represent the major improvement in the direct-photon analysis with respect to the previous publication~\cite{ALICE:2018gev}, in which the cocktail uncertainty reached up to 50\% in the $\eta$ meson mass range due to the absence of data at that time. Following the approach outlined in Ref.~\cite{ALICE:2020umb}, the \etapiz ratio is parameterised as a function of \pt using an empirical function~\cite{Altenkamper:2017qot}. Although less important for the direct-photon analysis, the input parameterisations of \phim~\cite{ALICE:2020jsh} and \jpsi~\cite{ALICE:2021dtt} have also been significantly improved based on the latest available data. As in the previous publication, \etap, \rhom, and \om mesons are generated assuming \mt scaling~\cite{Altenkamper:2017qot}, for which the particle yields are normalised at high \pt relative to the \piz yield as follows: 0.40 $\pm$ 0.08 for \etap (predicted by \textsc{Pythia}~6.4), 0.87 $\pm$ 0.17 for \rhom\cite{ALICE:2018qdv}, and \mbox{0.57 $\pm$ 0.11} for \om\cite{ALICE:2024YYYY}.

The charm and beauty production cross sections are evaluated using the dielectron spectrum in the intermediate mass range between \phim and \jpsi in the same way as reported in Ref.~\cite{ALICE:2018gev}, but using updated effective branching ratios of charm hadrons to dielectron pairs discussed in Ref.~\cite{ALICE:2024gyl}. The latter amounts to $7.33^{+0.61}_{-0.84}$\% for a charm quark to produce an electron after fragmentation and decay, which corresponds to $0.54 ^{+ 0.09}_{- 0.12}$\% for the $\rm{c}\rm{\bar{c}} \rightarrow \rm{e}^{+}\rm{e}^{-}$ case. The data are fitted simultaneously in \mee and \ptee with the templates of open charm and beauty production provided by \textsc{Pythia}~6.4~\cite{Sjostrand:2006za}, while keeping the contribution from LF and \jpsi decays fixed. The resulting cross sections amount to d$\sigma_{{\rm c\bar{c}}}$/d$y|_{y=0}$ = 1689 $\pm$ 124 (stat.) $\pm$ 152 (syst.) $^{+299}_{-296}$ (BR) $\mu$b and d$\sigma_{{\rm b\bar{b}}}$/d$y|_{y=0}$ = 82 $\pm$ 7 (stat.) $\pm$ 5 (syst.) $\pm$ 5 (BR) $\mu$b. These results supersede previously reported measurements~\cite{ALICE:2018gev} and are in agreement with them within uncertainties. The main difference in the uncertainties and central values comes from the newly available effective branching ratios as described in Ref.~\cite{ALICE:2024gyl} and better statistical and systematic uncertainties of the experimental data. These cross sections are consistent with independent measurements using hadronic decays~\cite{ALICE:2024gyl}.

For the high-multiplicity cocktail, the contribution from $\phi$ mesons is adjusted based on the multiplicity-dependent analysis described in Ref.~\cite{ALICE:2019etb}. As reported in Ref.~\cite{ALICE:2018gev}, contributions from \jpsi and correlated semi-leptonic heavy-flavour decays are weighted by a \pt-dependent multiplicity scaling factor based on the corresponding measurements as a function of charged-particle multiplicity in pp collisions at $\sqrt{s} = 13~{\rm TeV}$~\cite{ALICE:2020msa} and 7 TeV~\cite{ALICE:2015ikl}, respectively.

The following sources of systematic uncertainties are considered for the hadronic cocktail: the input parametrisations of the measured spectra as a function of \pt (\piz, $\eta$, $\phi$, \jpsi), the branching ratios of all included decay modes, the \mt scaling parameters, the resolution smearing, and the multiplicity scaling (for the HM analysis).
The systematic uncertainties of the hadronic cocktail in the invariant-mass region dominated by $\eta$ meson decays ($0.14 < \mee < 0.35$ \GeVmass) for \pt $>$ 1 \GeVc, where we extract the direct-photon signal, are 9.8--11.9$\%$ for INEL and 9.5--15.0$\%$ for HM.
\begin{figure}
  \centering     
  \includegraphics[width = 0.49\textwidth]{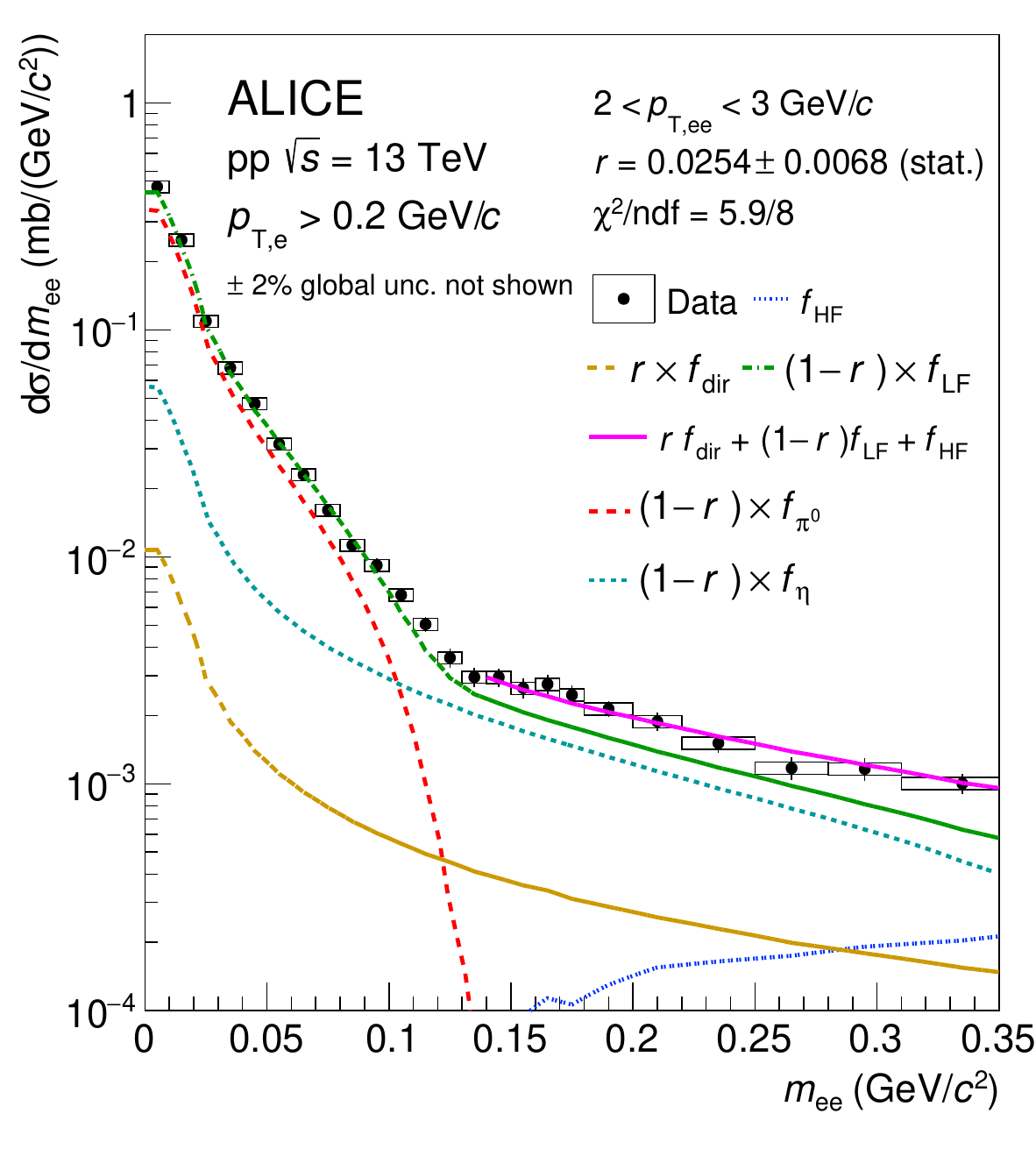}
  \includegraphics[width = 0.49\textwidth]{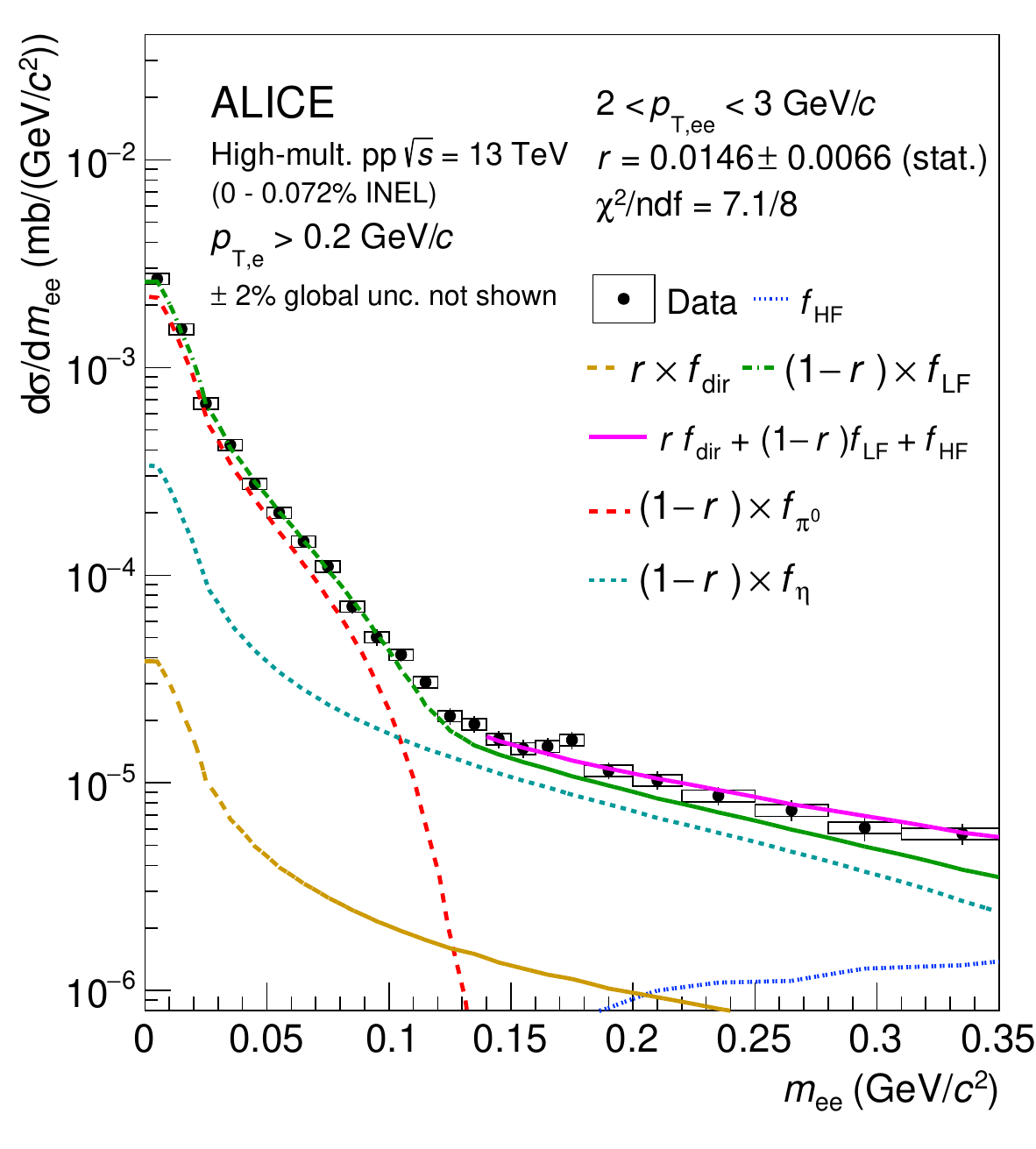}
  \caption{Fit of the dielectron cross section as a function of mass in INEL (left) and HM (right) pp collisions in the $2 < \ptee < 3$ \GeVc interval with a three-component fit function to extract the fraction of direct photons to inclusive photons $r$. Statistical and systematic uncertainties of the data are displayed as vertical bars and boxes, respectively. The integrated luminosities of \mbox{\ensuremath{\mathcal{L}^{\rm {MB}}_{\rm{int}}}\xspace = $29.8 \pm 0.6$ \nbarn} and \ensuremath{\mathcal{L}^{\rm {HM}}_{\rm{int}}}\xspace = $5.8 \pm 0.1$ \pbarn (2\% global uncertainty not shown in the Figure) are used to calculate the dielectron cross sections $\rm{d}\sigma/\rm{d}m_{\rm{ee}}$ from the differential dielectron yield per event. The resulting fraction of direct photons $r$ is shown together with its statistical uncertainty. The templates for dielectrons from LF, HF and direct photons are shown as different dashed lines, and the resulting fit is shown as a solid magenta line. The contributions from $\pi^{0}$ and $\eta$ mesons are also shown as separate lines. See the Appendix~\ref{sec:supplemental} of this letter for the fits of the dielectron cross section in other pair transverse momentum intervals (Figs.~\ref{fig:r_1_2}~\&~\ref{fig:r_3_6}).}
  \label{fig:r}
\end{figure}

Following the approach described in Refs.~\cite{ALICE:2018fvj,ALICE:2018ael}, the fraction of real direct photons over inclusive photons $r$ is extracted by a fit to the dielectron spectrum at invariant mass \mee $\ll$ \ptee assuming that this fraction is equal for virtual and real photons: $r = \gamma_{\rm dir}/\gamma_{\rm incl} = (\gamma^{*}_{\rm dir}/\gamma^{*}_{\rm incl})_{m_{{\rm ee}}\rightarrow0}$. The dielectron cross section is fitted in the mass range of $0.14 < \mee < 0.35$ \GeVmass with the following function as it is shown in Fig.~\hyperref[fig:r]{\ref{fig:r}}: $d\sigma/dm_{{\rm ee}} = r f_{\rm dir}(m_{{\rm ee}}) + (1-r)f_{\rm LF}(m_{{\rm ee}}) + f_{\rm HF}(m_{{\rm ee}})$. Here, $f_{\rm LF}$ and $f_{\rm HF}$ are the contributions from LF and HF decays, respectively, $f_{\rm dir}$ is the shape of virtual direct photon component determined by Kroll-Wada formula~\cite{Kroll:1955zu,Landsberg:1985gaz}, and the direct-photon fraction $r$ is the only fit parameter. The $f_{\rm HF}$ term is fixed to the extracted open charm and beauty cross sections at midrapidity.

Several sources of systematic uncertainties are considered for the direct-photon extraction that are summarised in Tables~\ref{tab:r_syst_inel} and~\ref{tab:r_syst_hm} of the Appendix~\ref{sec:supplemental}. The input experimental data on the dielectron cross sections is varied within its systematic uncertainties and rebinned using different mass binning. The dielectron mass range used to normalise $f_{\rm LF}$ and $f_{\rm dir}$ contributions to the data is changed between 0--20 \MeVmass and 0--40 \MeVmass limits, and the lower and upper edges of the fitting mass range are varied between 120--150 \MeVmass and 280--340 \MeVmass, correspondingly. The contribution from HF decays $f_{\rm HF}$ is varied within the corresponding uncertainties for measured cross sections d$\sigma_{{\rm c\bar{c}}}$/d$y|_{y=0}$ and d$\sigma_{{\rm b\bar{b}}}$/d$y|_{y=0}$, and an additional uncertainty related to the multiplicity scaling of HF production~\cite{ALICE:2015ikl} is taken into account for HM analysis. The dominant source of systematic uncertainty on the direct-photon fraction $r$ comes from the $\eta$ cocktail parameterisation for both INEL and HM cases. Thanks to the latest measurements of $\pi^{0}$ and $\eta$ mesons production in the same pp collision classes at $\sqrt{s} = 13$ TeV~\cite{ALICE:2024vgi}, the uncertainty on $\eta / \pi^{0}$ ratio amounts to 2--3\% in the relevant mass range. This value is significantly lower than in the previous analysis~\cite{ALICE:2018gev}, where the full difference between the $\eta$ meson yield estimate based on scarce data and $m_{\rm T}$ scaling was used to estimate the corresponding uncertainty. The total systematic uncertainty of up to 35\% (48\%) in INEL (HM) on the direct-photon fraction $r$ is obtained by summing in quadrature all individual contributions.

The direct photon yield $\gamma_{\rm{dir}}$ can be calculated from the yield of hadronic decay photons $\gamma_{\rm{decay}}$ using the fraction of direct over inclusive photons $r$ as:
\begin{equation}
\gamma_{\rm{dir}} = \gamma_{\rm{decay}} \times r / (1 - r).
\label{eq:gammadir}
\end{equation}
For this purpose, the spectra of real photons from hadronic decays are obtained by employing fast MC simulation of LF hadrons decaying into photons such as \piz, $\eta$, \etap, \rhom, \om and $\phi$. For each source, the input parameterisations of the corresponding \pt spectra are identical to those used in the dielectron cocktail simulation described above. Following the approach outlined in Ref.~\cite{ALICE:2018mjj}, mother particles are generated in the range of 0 $<$ \pt $<$ 50 \GeVc in the full azimuthal angle and are decayed into photons using \textsc{Pythia}~8. As for the dielectron sources, the dominant contribution to the decay-photon yield originates from \piz and $\eta$ mesons with a relative yield of about $\sim 88\%$ and $\sim 9\%$ at \pt $= 1$ \GeVc. The composition of the HM cocktail is similar to the one from the INEL. To estimate the systematic uncertainty of the decay-photon simulation, all particle parameterisations are changed within the uncertainties of their \pt spectra. This variation results in a maximum total uncertainty of 4\% on the real decay photon yield, with the dominant contribution from \piz decays.

\begin{figure}
  \centering
  \includegraphics[width = 0.47\textwidth]{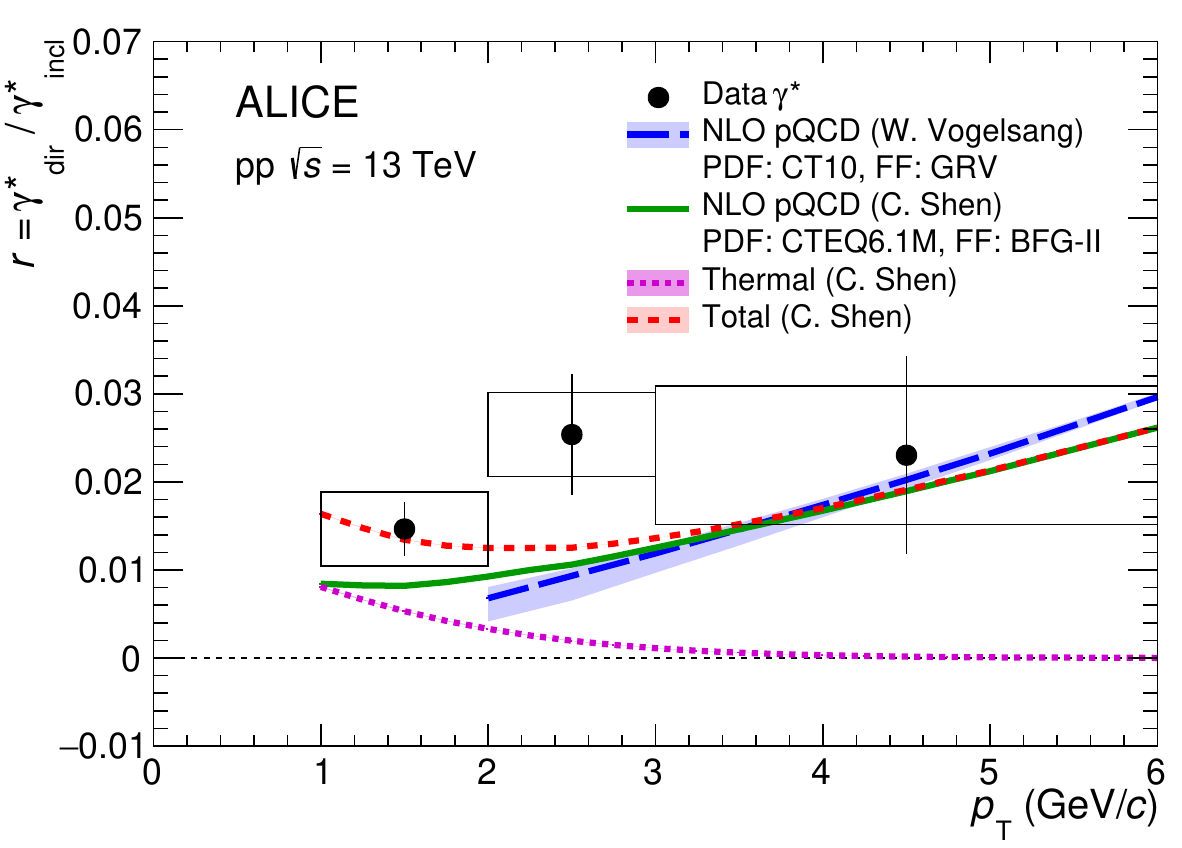}
  \includegraphics[width = 0.47\textwidth]{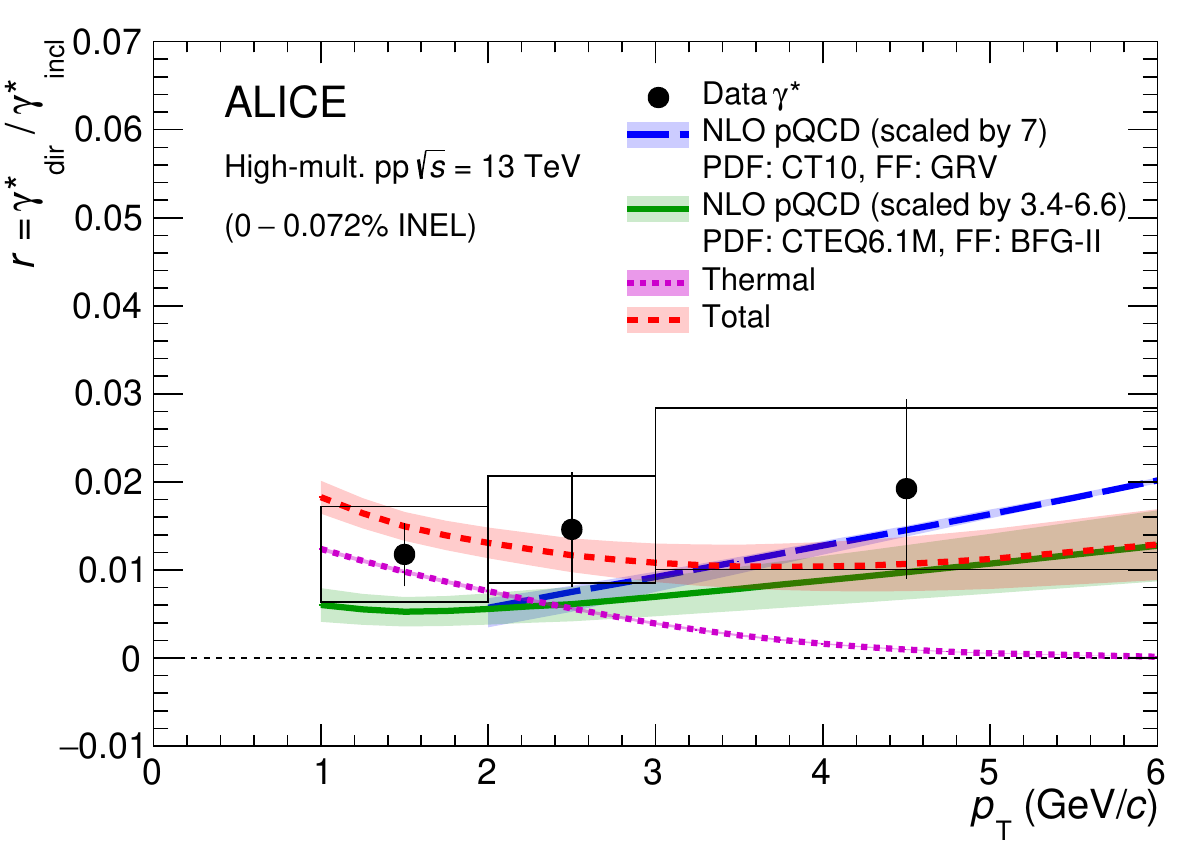}
  \caption{Direct photon fraction $r$ as a function of \pt extracted from fits to dielectron spectra in INEL (left) and HM (right) pp collisions. Statistical and systematic uncertainties of the data are displayed as vertical bars and boxes, respectively. The results in INEL pp collisions are compared to theoretical calculations of Refs.~\cite{Gordon:1993qc} and~\cite{Shen:2016zpp}. The results from HM pp data are compared to theoretical calculations of Ref.~\cite{Gordon:1993qc} for INEL pp collisions scaled by an empirical factor 7 based on \textsc{Pythia}~8 calculations (blue band). The red and green bands show the calculations of Ref.~\cite{Shen:2016zpp} scaled by the factor of 3.4--6.6, with the width of the bands representing the range of the applied scaling factor (see the text for more details).}
  \label{fig:r_vs_pt}
\end{figure}

\section{Results}
\label{sec:results}
Figure~\hyperref[fig:r_vs_pt]{\ref{fig:r_vs_pt}} shows the fraction of direct over inclusive photons $r$ as a function of \pt, which is found to be compatible between the two multiplicity classes. For the first time in pp collisions at LHC energies, experimental data show a direct-photon signal with 3.2$\sigma$ and 1.9$\sigma$ significance w.r.t.~null hypothesis (i.e.~$r = 0$ case) in the low-momentum range of $1 < \pt < 6$ \GeVc in INEL (left) and HM (right) event classes, respectively. These values are calculated using a Bayesian approach, for which statistical uncertainties are assumed to be point-to-point uncorrelated, whereas systematic uncertainties are treated as fully correlated. 

The results in INEL pp collisions are compared to pQCD calculations, which provide the direct photon spectra and can be employed to determine the direct-photon signal using the experimental decay photon simulation. The next-to-leading order (NLO) pQCD calculations by Vogelsang et al.~\cite{Gordon:1993qc} employ the CT10 parton distribution function~\cite{Lai:2010vv,Gao:2013xoa,Guzzi:2011sv} and the GRV fragmentation function described in Ref.~\cite{Gluck:1992zx}. The uncertainty band of the calculation from Ref.~\cite{Jager:2002xm} is given by the simultaneous variation of the factorisation, renormalisation, and fragmentation scales $\mu$ (0.5\pt $<$ $\mu$ $<$ 2\pt). The model of Shen et al.~\cite{Shen:2016zpp} uses the CTEQ6.1M parton distribution function~\cite{Stump:2003yu} and the BFG-II fragmentation function from Ref.~\cite{Bourhis:1997yu} for prompt photons and employs the $\mu = \pt$ scale to extrapolate the calculations to the low \pt region. It also includes the calculation of the thermal photon fraction, shown as an orange line in Fig.~\hyperref[fig:r_vs_pt]{\ref{fig:r_vs_pt}} (left), which is based on a viscous hydrodynamic evolution of the system with a lattice QCD-based equation of state~\cite{Shen:2016zpp}. For cells with a temperature above 180 MeV, leading-order QGP radiation based on Ref.~\cite{Arnold:2001ms} is assumed, and for cells with a temperature between 155 and 180 MeV hadronic photon production rates are calculated taking into account meson-meson scattering~\cite{Turbide:2003si}, the many-body $\rho$ spectral function~\cite{Rapp:1997fs} and pion-pion bremsstrahlung~\cite{Rapp:1999qu,Liu:2006imd}. Both calculations described above are consistent with the data in INEL pp collision within uncertainties. 

As for the HM data, the only available theoretical calculations to describe the direct-photon signal are the dedicated calculations of the thermal-photon contribution in high-multiplicity pp collisions from Shen et al.~\cite{Shen:2016zpp}. The comparison between the HM data and the prompt-photon yield calculated for INEL pp events shows that the latter do not describe the experimental HM data, with the corresponding $p$-value being less than 0.1 for both pQCD models. It is, therefore, likely that the calculations need an additional source of direct photons and/or an increased yield of prompt pQCD photons compared to INEL events to properly describe the HM data. The best description of the HM results was obtained by scaling the pQCD calculations for INEL pp collisions by Shen by a factor between 3.4 (lower limit) and 6.6 (upper limit) independent of transverse momentum and by adding the thermal-photon contribution calculated for HM events as described in Ref.~\cite{Shen:2016zpp}. It is important to note that the obtained scaling factor is in agreement with both the increase in charged-particle multiplicity in HM events of 5.9 w.r.t.~the INEL events~\cite{ALICE:2018gev} and with a larger yield of prompt photons calculated for HM data using \textsc{Pythia}~8 simulations. In the latter case, the resulting scaling factor amounts to about 7.0 w.r.t.~INEL event class, by which we scale the Vogelsang's pQCD calculations in INEL pp events (independent of transverse momentum) to compare them with the HM data. Without thermal photon contribution, the best description of the HM data is obtained by scaling of the pQCD photons by a factor of about 12. Figure~\hyperref[fig:r_vs_pt]{\ref{fig:r_vs_pt}} (right) compares the HM data and the scaled theoretical calculations for INEL pp collisions from both models.

\begin{figure}
  \centering     
  \includegraphics[width = 0.47\textwidth]{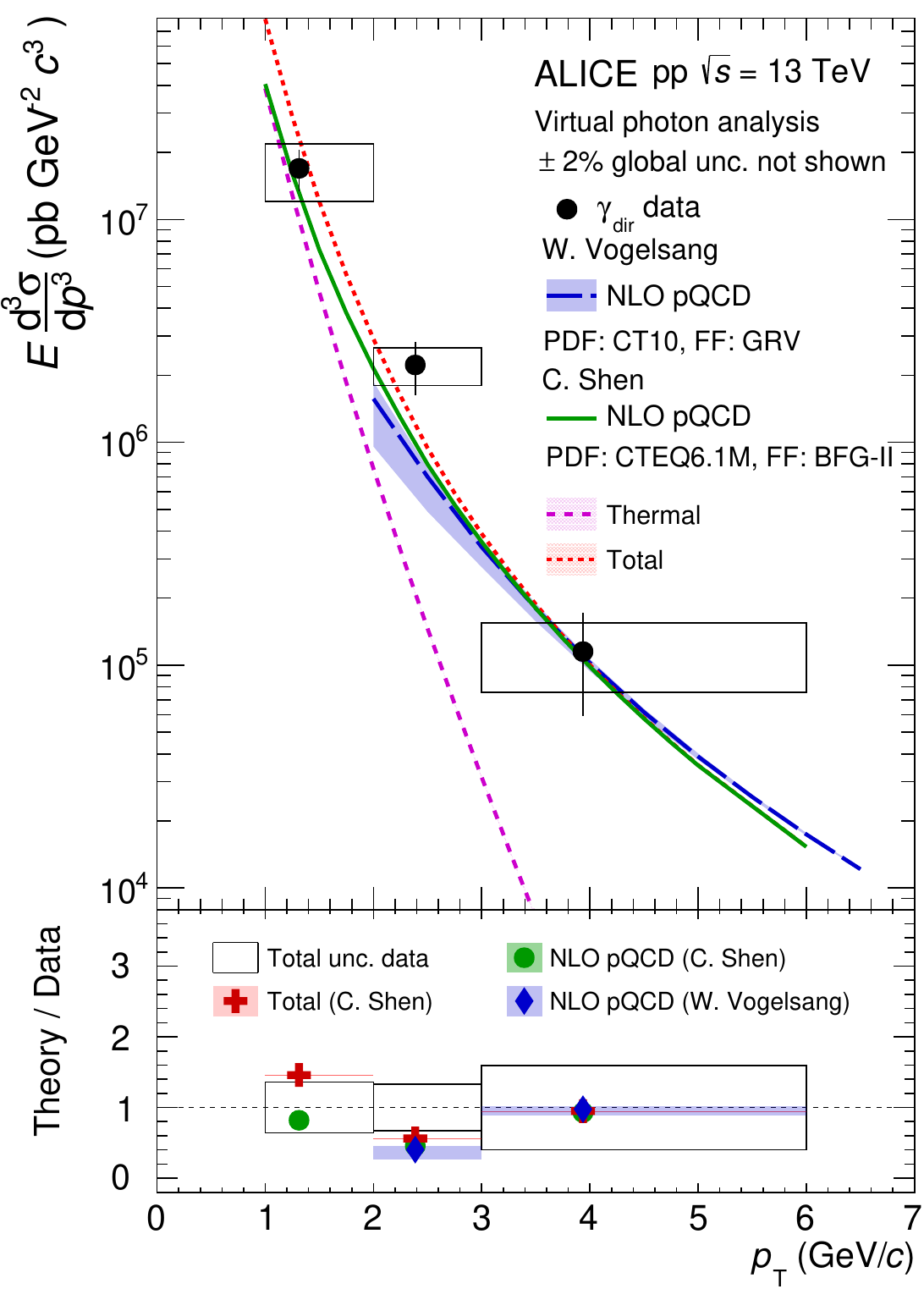}
  \includegraphics[width = 0.47\textwidth]{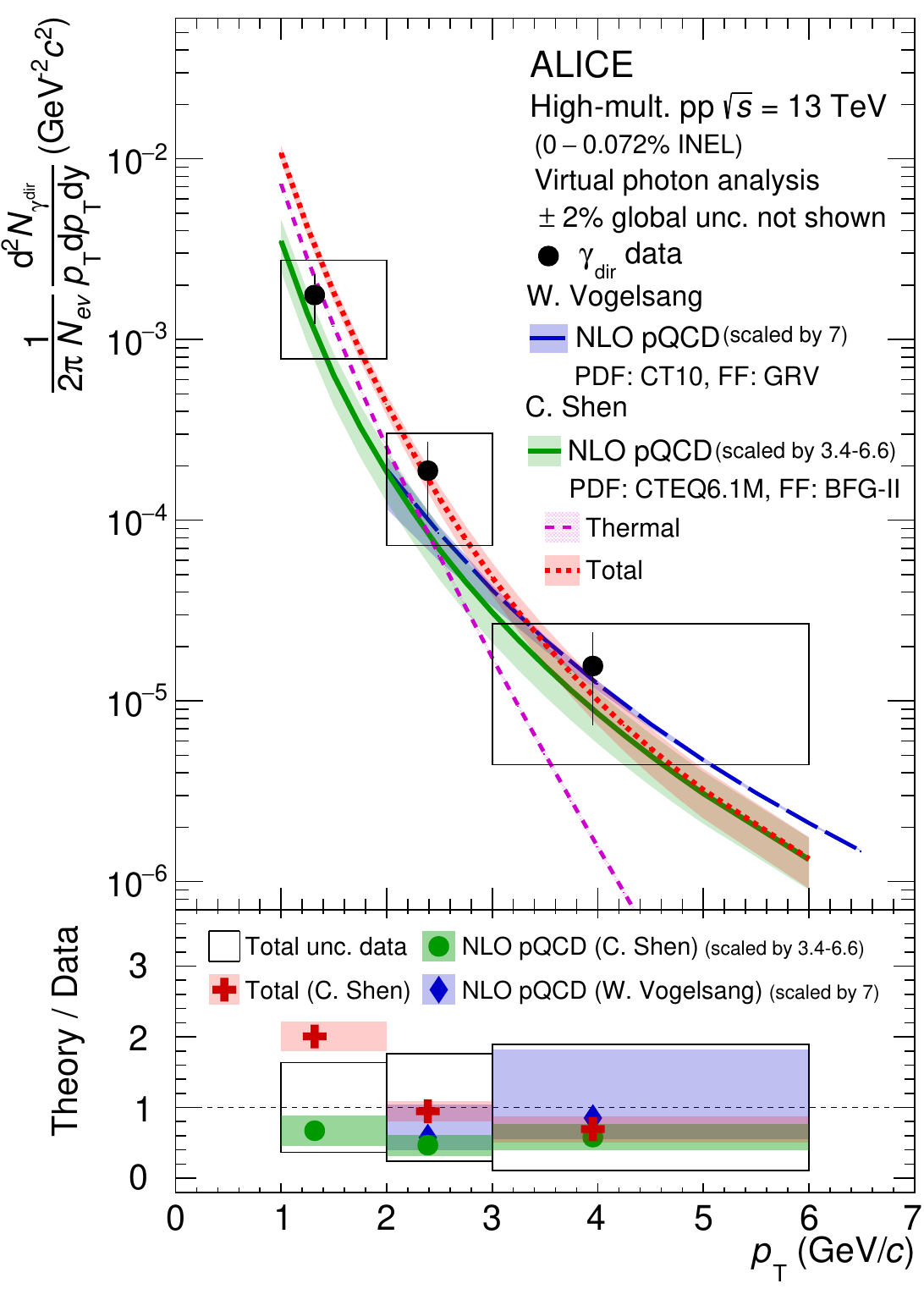}
  \caption{Direct-photon cross section (left) and invariant differential yield (right) as a function of \pt in INEL and HM pp collisions, respectively. Statistical and systematic uncertainties of the data are displayed as vertical bars and boxes, respectively. Global luminosity uncertainty of 2\% is not shown in the Figure. The results in INEL pp collisions are compared to theoretical calculations from~\cite{Gordon:1993qc} and~\cite{Shen:2016zpp}. The results from HM pp data are compared to theoretical calculations for INEL pp collisions scaled by an empirical factor of 3.4--6.6 (for Shen's calculations) and of 7 (for Vogelsang's model), see text for details. The uncertainty for Shen's pQCD calculations (green band) comes from the variation of the scaling factor, whereas the uncertainty of Vogelsang's pQCD results (blue band) is the original model uncertainty scaled by factor 7. The data points are plotted and the ratios are evaluated at the \pt values determined according to the Lafferty--Wyatt prescription~\cite{lafferty1995stick}, with the corresponding uncertainties being smaller than the marker size.}
  \label{fig:GammaPt}
\end{figure}

Figure~\hyperref[fig:GammaPt]{\ref{fig:GammaPt}} shows the direct-photon yield as a function of \pt obtained from the direct-photon fraction measurement following the Eq.~\ref{eq:gammadir}. Different lines on the left panel represent the same theoretical calculations which are also shown in Fig.~\hyperref[fig:r_vs_pt]{\ref{fig:r_vs_pt}} (left). The HM results on the right panel are compared to the theoretical calculations for INEL pp collisions scaled by empirical factors consistent with Fig.~\hyperref[fig:r_vs_pt]{\ref{fig:r_vs_pt}} (right).

\begin{figure}
  \centering     
  \includegraphics[width = 0.95\textwidth]{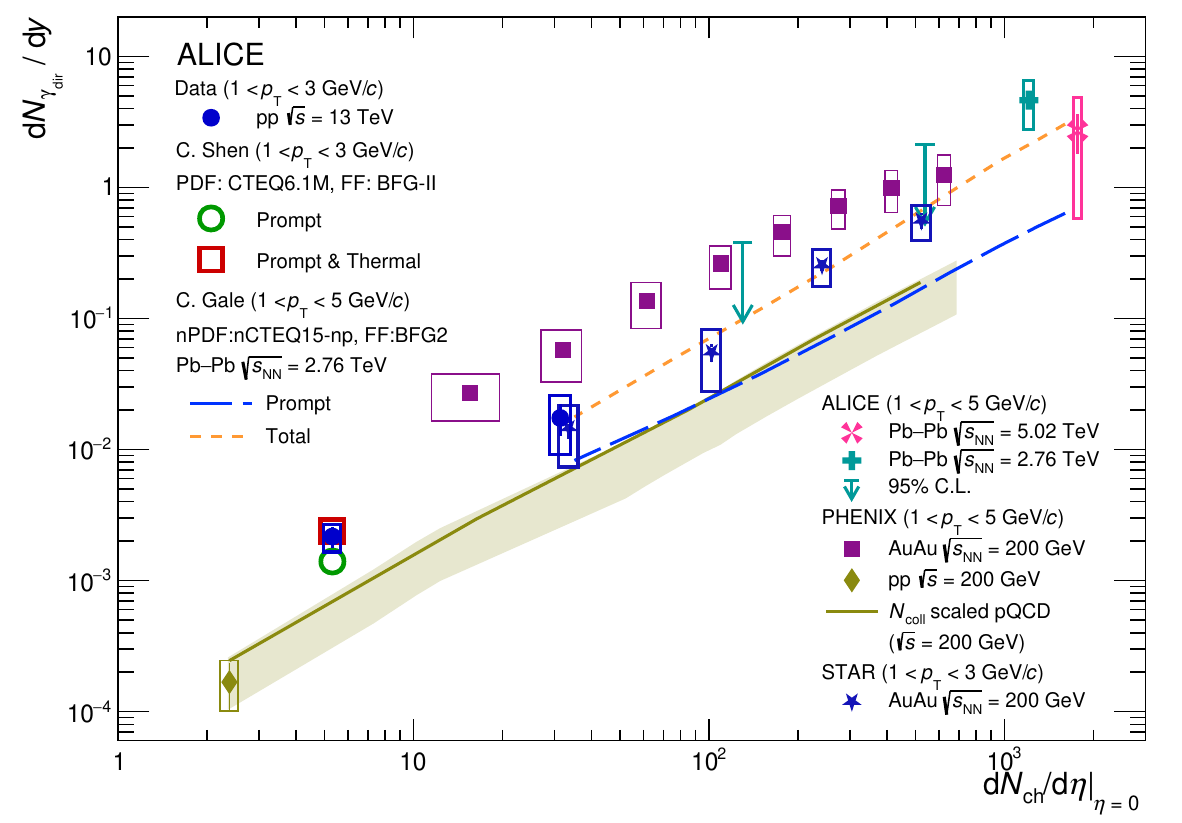}
  \caption{Direct photon yield at midrapidity as a function of charged-particle multiplicity \dndeta. The results from this letter are shown as two blue markers that correspond to the measurements in INEL and HM pp collisions at $\sqrt{s} = 13$ TeV, with the corresponding \dndeta values taken from Ref.~\cite{ALICE:2020swj}. The results in INEL event class are compared to theoretical calculations for prompt (open green circle) and prompt+thermal (open red square) yields~\cite{Shen:2016zpp}. The results from the STAR~\cite{STAR:2016use}, PHENIX~\cite{PHENIX:2018for,PHENIX:2022qfp} and ALICE~\cite{ALICE:2015xmh,ALICE:2023jef} collaborations are shown as full markers of various colours, and the band shows $N_{\rm coll}$-based extrapolation of pQCD calculations for pp collisions at $\sqrt{s} = 200$ GeV~\cite{PHENIX:2014nkk}. The \mbox{Pb--Pb} results are compared with the theoretical prediction by Gale~\cite{Gale:2021emg}, which includes prompt, pre-equilibrium and thermal photons. The prompt-photon yield is computed with NLO pQCD using the INCNLO code~\cite{Aurenche:1998gv} together with the nCTEQ15 nuclear parton distribution function~\cite{Kovarik:2015cma}, and the BFG-II~\cite{Bourhis:1997yu} fragmentation function.}
  \label{fig:GammaMult}
\end{figure}

Finally, Fig.~\hyperref[fig:GammaMult]{\ref{fig:GammaMult}} presents the integrated direct-photon yield as a function of charged-particle multiplicity at midrapidity measured in pp and heavy-ion collisions at RHIC~\cite{STAR:2016use,PHENIX:2018for,PHENIX:2022qfp} and LHC~\cite{ALICE:2015xmh,ALICE:2023jef} energies. For this purpose, direct-photon yields in the INEL and HM event classes presented in Fig.~\ref{fig:GammaPt} are integrated over the range 1 $<$ \pt $<$ 3 \GeVc. The corresponding \dndeta values represent the pseudorapidity density of charged particles at midrapidity $|\eta| < 0.5$ measured in the same multiplicity classes~\cite{ALICE:2020swj}. Whereas the direct-photon fraction $r$ is compatible for both multiplicity classes, a much higher yield of direct photons per event is observed in the HM class.
The results in the HM event class reach a multiplicity close to that in peripheral Au--Au collisions at \mbox{$\sqrt{s_{\rm NN}} = 200$~GeV}~\cite{PHENIX:2022qfp} and show a similar integrated yield of direct photons. The results are compatible with the data reported by the STAR collaboration in Ref.~\cite{STAR:2016use} at similar \dndeta, which on the one hand show a significantly lower direct photon yield compared to the results from the PHENIX collaboration presented in Ref.~\cite{PHENIX:2018for} at the same multiplicity, but on the other hand hint at an excess of direct photons with respect to~$N_{\rm coll}$-based extrapolation from pp collisions at $\sqrt{s} =$ 200 GeV~\cite{PHENIX:2014nkk}. Following the approach described in Refs.~\cite{PHENIX:2022rsx,PHENIX:2022qfp}, the direct-photon yield reported in this letter is tested against an assumption of a universal scaling behaviour with charged-particle multiplicity as $(\dndeta)^{\alpha}$. The fit to the newly available pp data from ALICE leads to \mbox{$\alpha = 1.17 \pm 0.18 \rm{(stat.)} \pm 0.21\rm{(syst.)}$}, which is in agreement with the values previously reported by the PHENIX Collaboration for Au--Au data~\cite{PHENIX:2022rsx,PHENIX:2022qfp}.

In summary, in this letter we have presented direct-photon measurements in inelastic and high-multiplicity pp collisions at $\sqrt{s} = 13$ TeV. With increased statistics and significantly reduced systematic uncertainties of the hadronic cocktail compared to the previous publication~\cite{ALICE:2018gev}, we measure for the first time a significant direct photon yield at low transverse momentum in pp collisions at the LHC energies.
The direct photon signal is seen in both event classes that were studied. The results in the inelastic event class are found to be consistent with pQCD-based calculations, while the lack of theoretical predictions prevents a conclusion on the possible presence (and yield) of thermal radiation in high-multiplicity pp collisions. However, from simple scaling of pQCD-based calculations for inelastic pp collisions, it was found that the best description of high-multiplicity data is obtained with the sum of thermal photons (calculated for the HM event class as described in Ref.~\cite{Shen:2016zpp}) and the prompt-photon contribution scaled by a factor of 3.4--6.6 from inelastic pp collisions. The large uncertainty in the scaling factor reflects the still sizeable uncertainties in the currently available experimental high-multiplicity data. The multiplicity range covered in the present analysis is of particular interest in the context of the search for a possible onset of thermal radiation in small systems. The statistics that will be collected during the Run 3 and Run 4 periods of LHC operation will allow one to study the direct photon yield with higher statistical precision and in a more differential way. The results presented in this Letter and the future results from Run 3 data  should stimulate the interest of the theoretical community to calculate the yield of direct photons in small systems as a function of charged-particle multiplicity to shed light on the possible presence of a QGP in small systems.


\newenvironment{acknowledgement}{\relax}{\relax}
\begin{acknowledgement}
\section*{Acknowledgements}
We would like to thank Werner Vogelsang and Chun Shen for providing the theoretical predictions shown in this paper. 

The ALICE Collaboration would like to thank all its engineers and technicians for their invaluable contributions to the construction of the experiment and the CERN accelerator teams for the outstanding performance of the LHC complex.
The ALICE Collaboration gratefully acknowledges the resources and support provided by all Grid centres and the Worldwide LHC Computing Grid (WLCG) collaboration.
The ALICE Collaboration acknowledges the following funding agencies for their support in building and running the ALICE detector:
A. I. Alikhanyan National Science Laboratory (Yerevan Physics Institute) Foundation (ANSL), State Committee of Science and World Federation of Scientists (WFS), Armenia;
Austrian Academy of Sciences, Austrian Science Fund (FWF): [M 2467-N36] and Nationalstiftung f\"{u}r Forschung, Technologie und Entwicklung, Austria;
Ministry of Communications and High Technologies, National Nuclear Research Center, Azerbaijan;
Conselho Nacional de Desenvolvimento Cient\'{\i}fico e Tecnol\'{o}gico (CNPq), Financiadora de Estudos e Projetos (Finep), Funda\c{c}\~{a}o de Amparo \`{a} Pesquisa do Estado de S\~{a}o Paulo (FAPESP) and Universidade Federal do Rio Grande do Sul (UFRGS), Brazil;
Bulgarian Ministry of Education and Science, within the National Roadmap for Research Infrastructures 2020-2027 (object CERN), Bulgaria;
Ministry of Education of China (MOEC) , Ministry of Science \& Technology of China (MSTC) and National Natural Science Foundation of China (NSFC), China;
Ministry of Science and Education and Croatian Science Foundation, Croatia;
Centro de Aplicaciones Tecnol\'{o}gicas y Desarrollo Nuclear (CEADEN), Cubaenerg\'{\i}a, Cuba;
Ministry of Education, Youth and Sports of the Czech Republic, Czech Republic;
The Danish Council for Independent Research | Natural Sciences, the VILLUM FONDEN and Danish National Research Foundation (DNRF), Denmark;
Helsinki Institute of Physics (HIP), Finland;
Commissariat \`{a} l'Energie Atomique (CEA) and Institut National de Physique Nucl\'{e}aire et de Physique des Particules (IN2P3) and Centre National de la Recherche Scientifique (CNRS), France;
Bundesministerium f\"{u}r Bildung und Forschung (BMBF) and GSI Helmholtzzentrum f\"{u}r Schwerionenforschung GmbH, Germany;
General Secretariat for Research and Technology, Ministry of Education, Research and Religions, Greece;
National Research, Development and Innovation Office, Hungary;
Department of Atomic Energy Government of India (DAE), Department of Science and Technology, Government of India (DST), University Grants Commission, Government of India (UGC) and Council of Scientific and Industrial Research (CSIR), India;
National Research and Innovation Agency - BRIN, Indonesia;
Istituto Nazionale di Fisica Nucleare (INFN), Italy;
Japanese Ministry of Education, Culture, Sports, Science and Technology (MEXT) and Japan Society for the Promotion of Science (JSPS) KAKENHI, Japan;
Consejo Nacional de Ciencia (CONACYT) y Tecnolog\'{i}a, through Fondo de Cooperaci\'{o}n Internacional en Ciencia y Tecnolog\'{i}a (FONCICYT) and Direcci\'{o}n General de Asuntos del Personal Academico (DGAPA), Mexico;
Nederlandse Organisatie voor Wetenschappelijk Onderzoek (NWO), Netherlands;
The Research Council of Norway, Norway;
Pontificia Universidad Cat\'{o}lica del Per\'{u}, Peru;
Ministry of Science and Higher Education, National Science Centre and WUT ID-UB, Poland;
Korea Institute of Science and Technology Information and National Research Foundation of Korea (NRF), Republic of Korea;
Ministry of Education and Scientific Research, Institute of Atomic Physics, Ministry of Research and Innovation and Institute of Atomic Physics and Universitatea Nationala de Stiinta si Tehnologie Politehnica Bucuresti, Romania;
Ministry of Education, Science, Research and Sport of the Slovak Republic, Slovakia;
National Research Foundation of South Africa, South Africa;
Swedish Research Council (VR) and Knut \& Alice Wallenberg Foundation (KAW), Sweden;
European Organization for Nuclear Research, Switzerland;
Suranaree University of Technology (SUT), National Science and Technology Development Agency (NSTDA) and National Science, Research and Innovation Fund (NSRF via PMU-B B05F650021), Thailand;
Turkish Energy, Nuclear and Mineral Research Agency (TENMAK), Turkey;
National Academy of  Sciences of Ukraine, Ukraine;
Science and Technology Facilities Council (STFC), United Kingdom;
National Science Foundation of the United States of America (NSF) and United States Department of Energy, Office of Nuclear Physics (DOE NP), United States of America.
In addition, individual groups or members have received support from:
Czech Science Foundation (grant no. 23-07499S), Czech Republic;
FORTE project, reg.\ no.\ CZ.02.01.01/00/22\_008/0004632, Czech Republic, co-funded by the European Union, Czech Republic;
European Research Council (grant no. 950692), European Union;
ICSC - Centro Nazionale di Ricerca in High Performance Computing, Big Data and Quantum Computing, European Union - NextGenerationEU;
Academy of Finland (Center of Excellence in Quark Matter) (grant nos. 346327, 346328), Finland;
Deutsche Forschungs Gemeinschaft (DFG, German Research Foundation) ``Neutrinos and Dark Matter in Astro- and Particle Physics'' (grant no. SFB 1258), Germany.

\end{acknowledgement}

\bibliographystyle{utphys}   
\bibliography{bibliography}

\newpage
\appendix

\section{Supplemental Material}
\label{sec:supplemental}

This Appendix reports the fits of the dielectron cross section as a function of mass in the pair transverse momentum intervals of $1 < \ptee < 2$ \GeVc and $3 < \ptee < 6$ \GeVc, in addition to the fits that are shown in Fig.~\ref{fig:r}, as well as the Tables that summarise different sources of systematic uncertainties relevant for the analysis.

\begin{figure}[ht]
  \centering     
  \includegraphics[width = 0.49\textwidth]{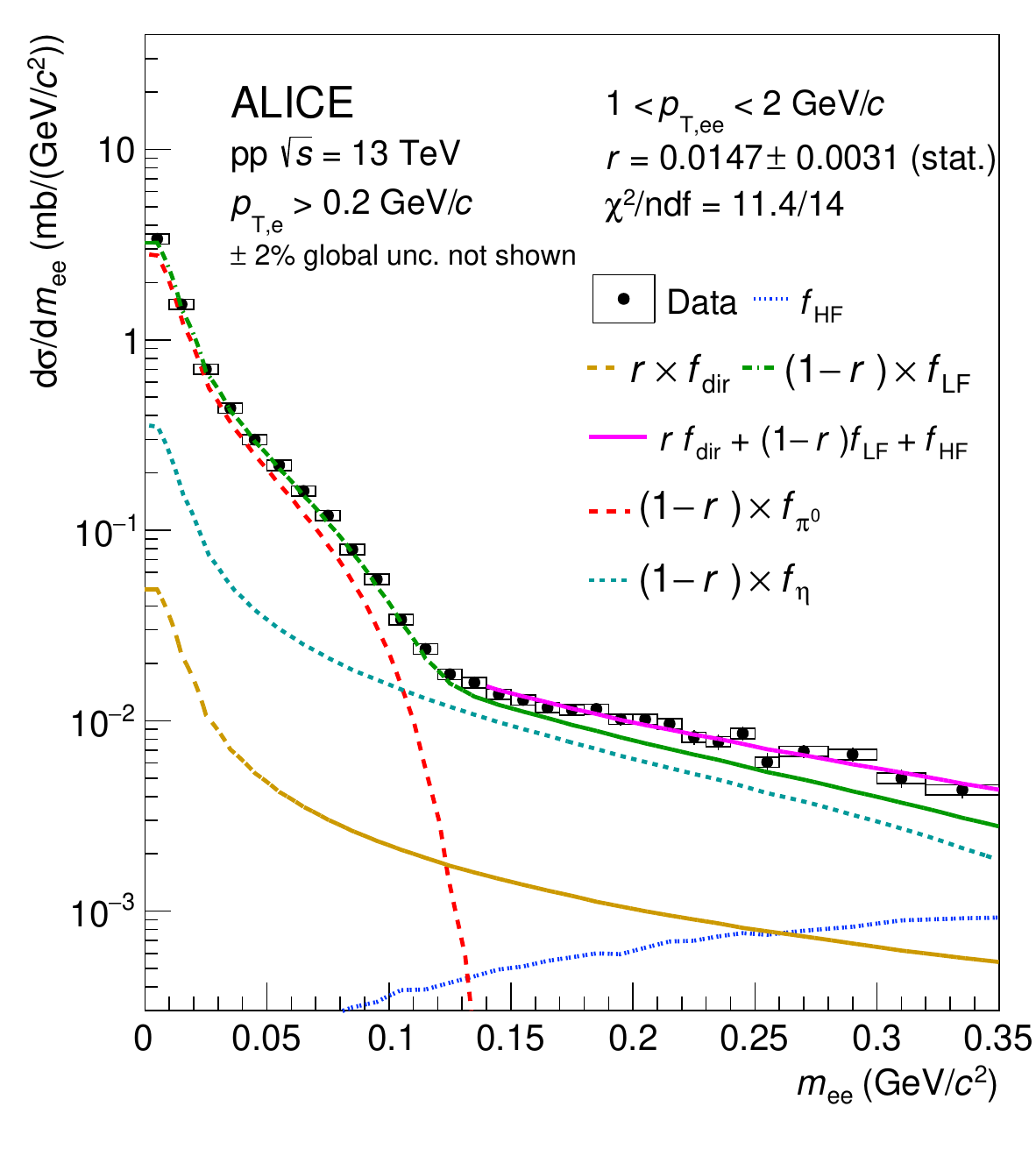}
  \includegraphics[width = 0.49\textwidth]{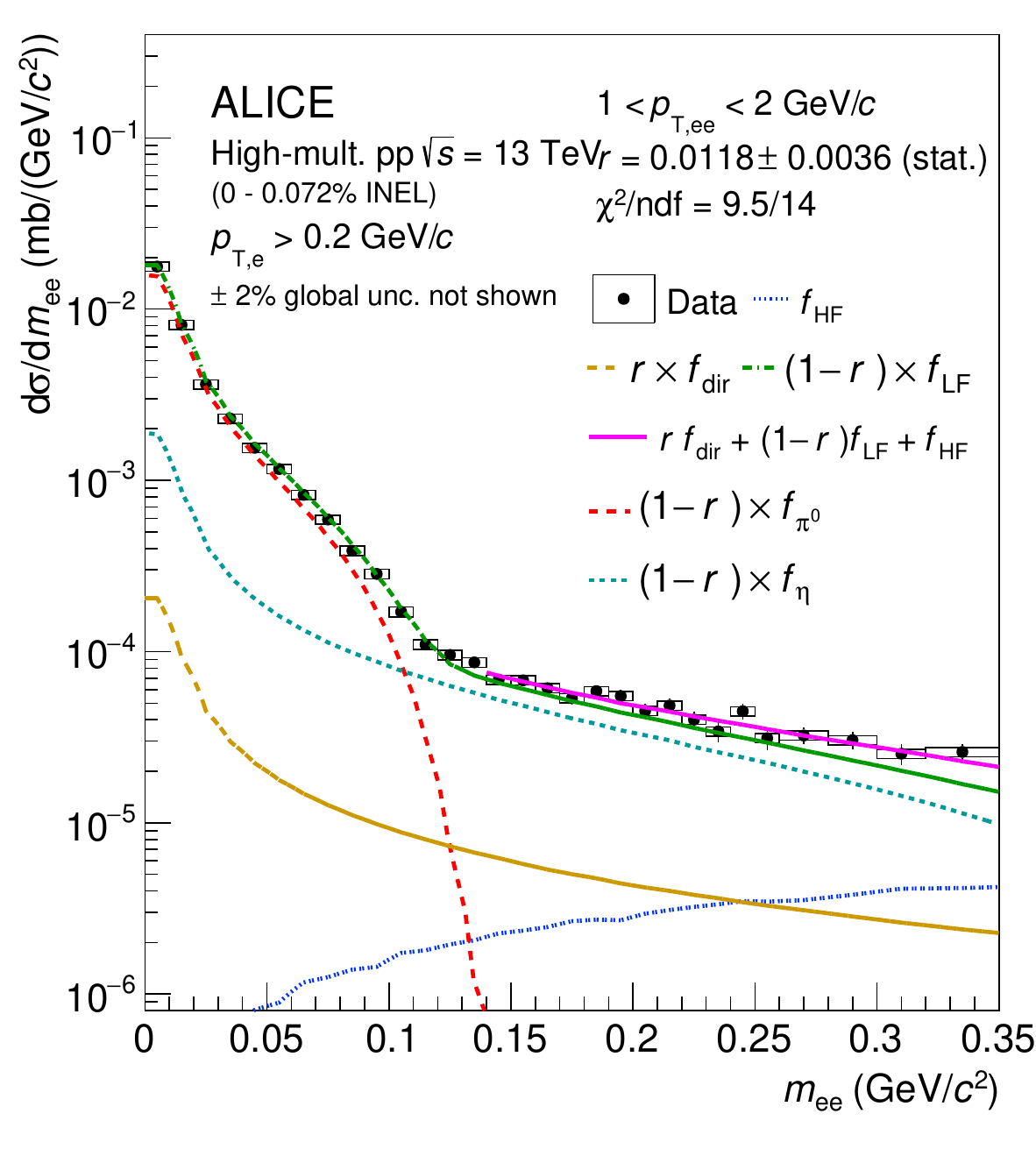}
  \caption{Fit of the dielectron cross section as a function of mass in INEL (left) and HM (right) pp collisions in the $1 < \ptee < 2$ \GeVc interval with a three-component fit function to extract the fraction of direct photons to inclusive photons $r$. Statistical and systematic uncertainties of the data are displayed as vertical bars and boxes, respectively. Global luminosity uncertainty of 2\% is not shown in the Figure. The resulting fraction of direct photons $r$ is shown together with its statistical uncertainty. The templates for dielectrons from LF, HF and direct photons are shown as different dashed lines, and the resulting fit is shown as a solid magenta line. The contributions from $\pi^{0}$ and $\eta$ mesons are also shown as separate lines.}
  \label{fig:r_1_2}
\end{figure}

\begin{figure}[ht]
  \centering     
  \includegraphics[width = 0.49\textwidth]{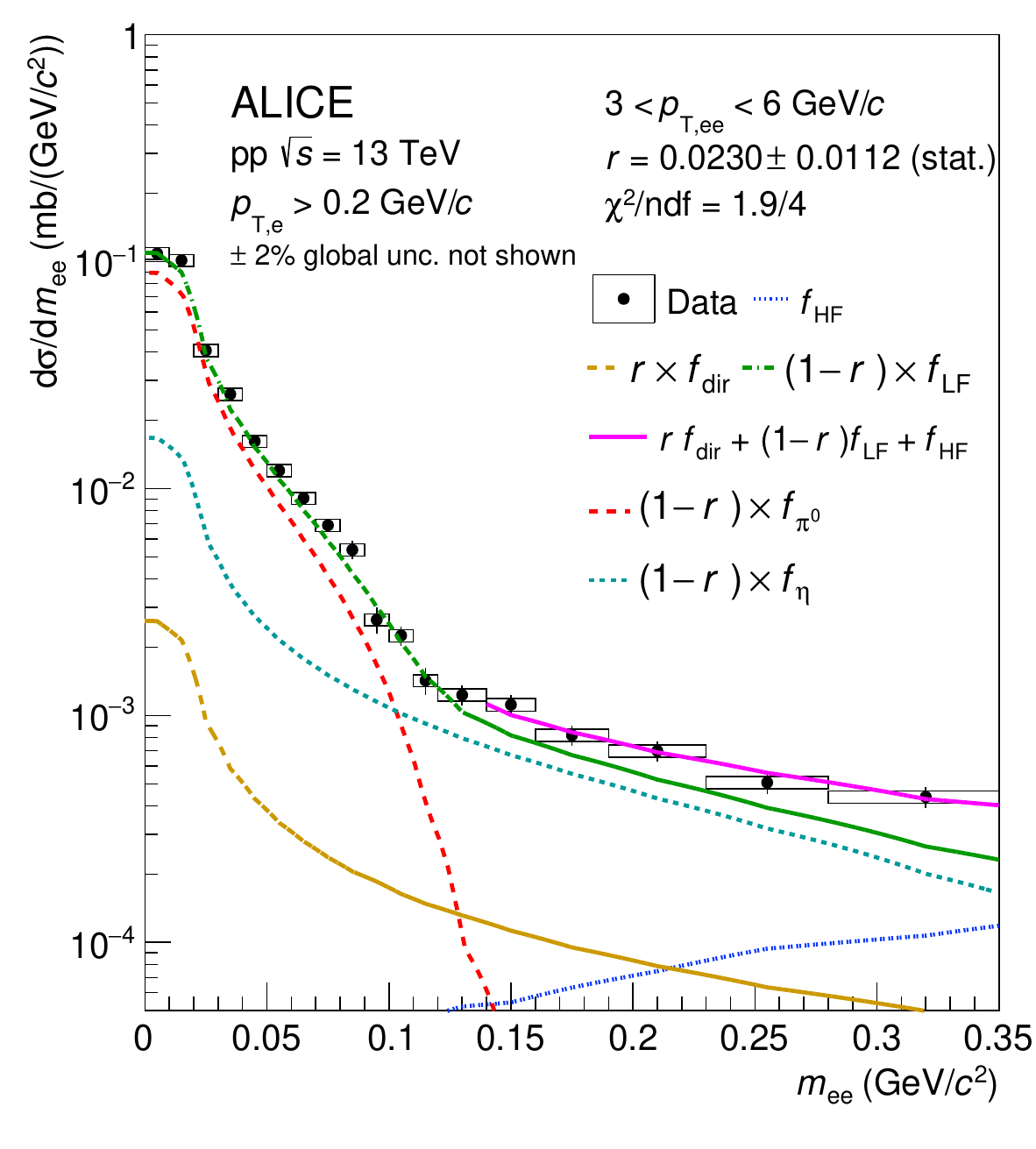}
  \includegraphics[width = 0.49\textwidth]{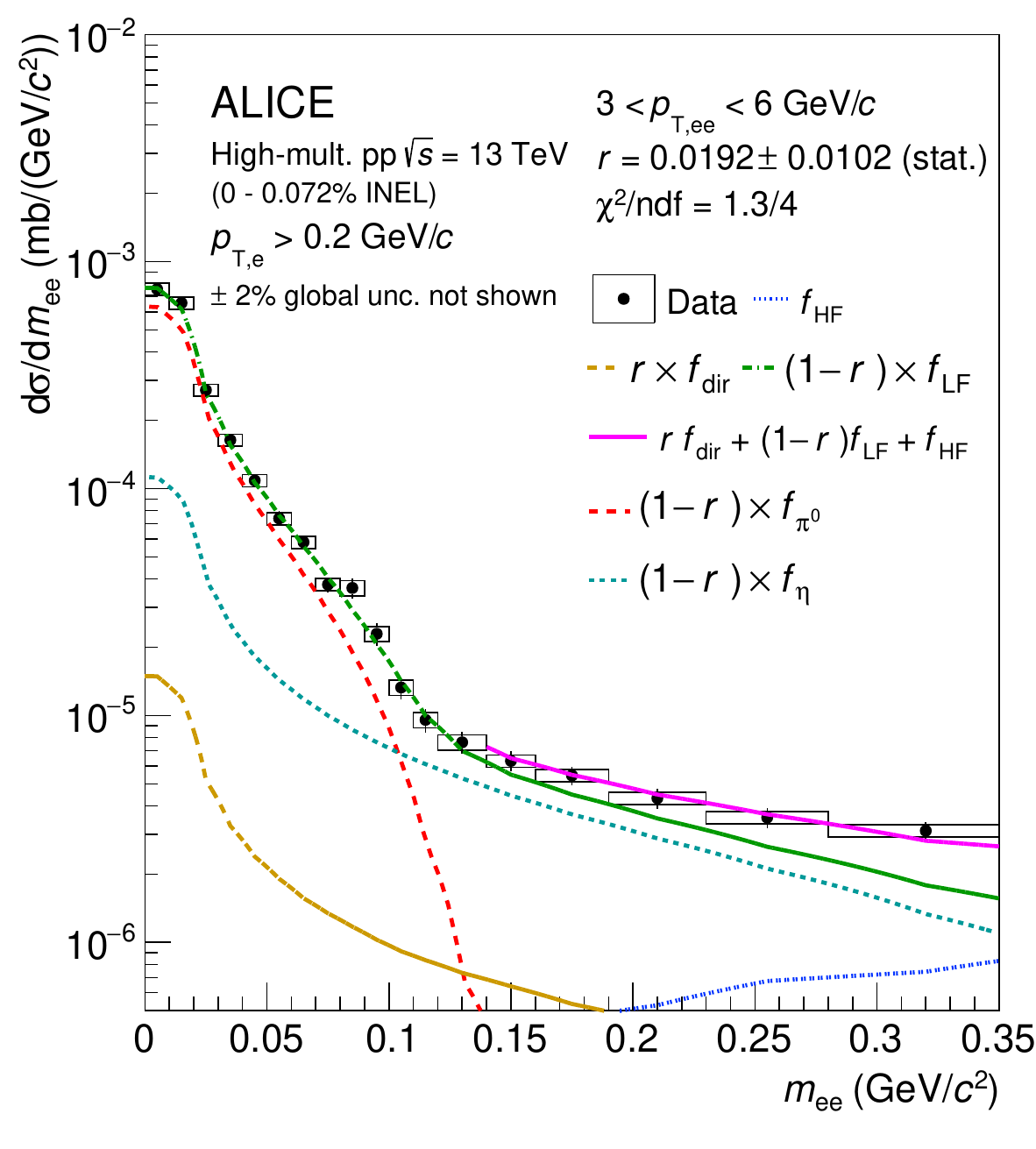}
  \caption{Fit of the dielectron cross section as a function of mass in INEL (left) and HM (right) pp collisions in the $3 < \ptee < 6$ \GeVc interval with a three-component fit function to extract the fraction of direct photons to inclusive photons $r$. Statistical and systematic uncertainties of the data are displayed as vertical bars and boxes, respectively. Global luminosity uncertainty of 2\% is not shown in the Figure. The resulting fraction of direct photons $r$ is shown together with its statistical uncertainty. The templates for dielectrons from LF, HF and direct photons are shown as different dashed lines, and the resulting fit is shown as a solid magenta line. The contributions from $\pi^{0}$ and $\eta$ mesons are also shown as separate lines.}
  \label{fig:r_3_6}
\end{figure}

\clearpage

\begin{table}[h]
    \centering
    \begin{tabular}{lcc}
        \toprule
        & INEL & HM \\
        \midrule
        Track reconstruction in the TPC and electron identification & 1--4.5$\%$ & 1--8$\%$  \\
        Track reconstruction in the ITS & 2\% & 3\% \\
        ITS--TPC matching & 6\% & 6\%  \\
        Photon conversion rejection ($m_{\rm{ee}} < 100$ \MeVmass) & 0.9--2.2$\%$ & 0.5--5.3$\%$  \\
        Acceptance correction factor $R$ & 2\% & 2\%  \\
        \midrule
        Total              & 6.5--8.1$\%$  & 6.8--12$\%$ \\
        \bottomrule
    \end{tabular}
    \caption{Overview of systematic uncertainties contributing to the measurement of dielectron cross section. The mass- and \ptee-dependent uncertainties are reported as ranges and tend to increase towards higher mass and \ptee.}
    \label{tab:mee_syst}
\end{table}

\begin{table}[h]
    \centering
    \begin{tabular}{lccc}
        \toprule
        \ptee interval (\GeVc) & 1--2 & 2--3 & 3--6 \\
        \midrule
        Input experimental data             & 7.5\%   & 6.5\%   & 25.4\%  \\
        Normalization range                 & 5.8\%   & 4.9\%   & 10.9\%    \\
        Fitting range                       & 6.7\% & 6.3\%  & 3.7\% \\
        Light-flavour cocktail              & 28\%   & 16.3\%   & 21.3\%    \\
        Heavy-flavour cocktail              & 3.6\% & 2.6\% & 4.2\% \\
        \midrule
        Total                               & 30.5\% & 19.5\% & 35.3\% \\
        \bottomrule
    \end{tabular}
    \caption{Summary of relative systematic uncertainties for the fraction of virtual direct photons $r$ in INEL events.}
    \label{tab:r_syst_inel}
\end{table}

\begin{table}[h]
    \centering
    \begin{tabular}{lccc}
        \toprule
        \ptee interval (\GeVc) & 1--2 & 2--3 & 3--6 \\
        \midrule
        Input experimental data                           & 26.9\%  & 23.6\%  & 33.8\%  \\
        Normalization range                               & 6.4\%  & 8.1\%  & 5.9\%  \\
        Fitting range                                     & 7.3\%  & 14\%  & 18\%  \\
        Light-flavour cocktail                            & 30.7\%  & 27.9\%  & 25\%  \\
        Heavy-flavour cocktail                            & 3.6\%  & 4.4\%  & 5\%  \\
        Mult. scaling of HF cocktail                      & 17.9\%  & 8.7\%  & 10.8\%  \\
        \midrule
        Total                                             & 45.8\%  & 41.1\%  & 47.8\%  \\
        \bottomrule
    \end{tabular}
    \caption{Summary of relative systematic uncertainties for the fraction of virtual direct photons $r$ in HM events.}    
    \label{tab:r_syst_hm}
\end{table}

\newpage
\section{The ALICE Collaboration}
\label{app:collab}
\begin{flushleft} 
\small

S.~Acharya\,\orcidlink{0000-0002-9213-5329}\,$^{\rm 126}$, 
A.~Agarwal$^{\rm 134}$, 
G.~Aglieri Rinella\,\orcidlink{0000-0002-9611-3696}\,$^{\rm 32}$, 
L.~Aglietta\,\orcidlink{0009-0003-0763-6802}\,$^{\rm 24}$, 
M.~Agnello\,\orcidlink{0000-0002-0760-5075}\,$^{\rm 29}$, 
N.~Agrawal\,\orcidlink{0000-0003-0348-9836}\,$^{\rm 25}$, 
Z.~Ahammed\,\orcidlink{0000-0001-5241-7412}\,$^{\rm 134}$, 
S.~Ahmad\,\orcidlink{0000-0003-0497-5705}\,$^{\rm 15}$, 
S.U.~Ahn\,\orcidlink{0000-0001-8847-489X}\,$^{\rm 71}$, 
I.~Ahuja\,\orcidlink{0000-0002-4417-1392}\,$^{\rm 36}$
% ...

\section*{Affiliation Notes}

$^{\rm I}$ Deceased\\
$^{\rm II}$ Also at: Max-Planck-Institut fur Physik, Munich, Germany\\
$^{\rm III}$ Also at: Italian National Agency for New Technologies, Energy and Sustainable Economic Development (ENEA), Bologna, Italy\\
$^{\rm IV}$ Also at: Dipartimento DET del Politecnico di Torino, Turin, Italy\\
$^{\rm V}$ Also at: Department of Applied Physics, Aligarh Muslim University, Aligarh, India\\
$^{\rm VI}$ Also at: Institute of Theoretical Physics, University of Wroclaw, Poland\\
$^{\rm VII}$ Also at: Facultad de Ciencias, Universidad Nacional Aut\'{o}noma de M\'{e}xico, Mexico City, Mexico\\
$^{\rm VIII}$ Also at: An institution covered by a cooperation agreement with CERN\\

\section*{Collaboration Institutes}

$^{1}$ A.I. Alikhanyan National Science Laboratory (Yerevan Physics Institute) Foundation, Yerevan, Armenia\\
$^{2}$ AGH University of Krakow, Cracow, Poland\\
$^{3}$ Bogolyubov Institute for Theoretical Physics, National Academy of Sciences of Ukraine, Kiev, Ukraine\\
$^{4}$ Bose Institute, Department of Physics  and Centre for Astroparticle Physics and Space Science (CAPSS), Kolkata, India\\
$^{5}$ California Polytechnic State University, San Luis Obispo, California, United States\\
$^{6}$ Central China Normal University, Wuhan, China\\
$^{7}$ Centro de Aplicaciones Tecnol\'{o}gicas y Desarrollo Nuclear (CEADEN), Havana, Cuba\\
$^{8}$ Centro de Investigaci\'{o}n y de Estudios Avanzados (CINVESTAV), Mexico City and M\'{e}rida, Mexico\\
$^{9}$ Chicago State University, Chicago, Illinois, United States\\
$^{10}$ China Nuclear Data Center, China Institute of Atomic Energy, Beijing, China\\
$^{11}$ China University of Geosciences, Wuhan, China\\
$^{12}$ Chungbuk National University, Cheongju, Republic of Korea\\
$^{13}$ Comenius University Bratislava, Faculty of Mathematics, Physics and Informatics, Bratislava, Slovak Republic\\
$^{14}$ Creighton University, Omaha, Nebraska, United States\\
$^{15}$ Department of Physics, Aligarh Muslim University, Aligarh, India\\
$^{16}$ Department of Physics, Pusan National University, Pusan, Republic of Korea\\
$^{17}$ Department of Physics, Sejong University, Seoul, Republic of Korea\\
$^{18}$ Department of Physics, University of California, Berkeley, California, United States\\
$^{19}$ Department of Physics, University of Oslo, Oslo, Norway\\
$^{20}$ Department of Physics and Technology, University of Bergen, Bergen, Norway\\
$^{21}$ Dipartimento di Fisica, Universit\`{a} di Pavia, Pavia, Italy\\
$^{22}$ Dipartimento di Fisica dell'Universit\`{a} and Sezione INFN, Cagliari, Italy\\
$^{23}$ Dipartimento di Fisica dell'Universit\`{a} and Sezione INFN, Trieste, Italy\\
$^{24}$ Dipartimento di Fisica dell'Universit\`{a} and Sezione INFN, Turin, Italy\\
$^{25}$ Dipartimento di Fisica e Astronomia dell'Universit\`{a} and Sezione INFN, Bologna, Italy\\
$^{26}$ Dipartimento di Fisica e Astronomia dell'Universit\`{a} and Sezione INFN, Catania, Italy\\
$^{27}$ Dipartimento di Fisica e Astronomia dell'Universit\`{a} and Sezione INFN, Padova, Italy\\
$^{28}$ Dipartimento di Fisica `E.R.~Caianiello' dell'Universit\`{a} and Gruppo Collegato INFN, Salerno, Italy\\
$^{29}$ Dipartimento DISAT del Politecnico and Sezione INFN, Turin, Italy\\
$^{30}$ Dipartimento di Scienze MIFT, Universit\`{a} di Messina, Messina, Italy\\
$^{31}$ Dipartimento Interateneo di Fisica `M.~Merlin' and Sezione INFN, Bari, Italy\\
$^{32}$ European Organization for Nuclear Research (CERN), Geneva, Switzerland\\
$^{33}$ Faculty of Electrical Engineering, Mechanical Engineering and Naval Architecture, University of Split, Split, Croatia\\
$^{34}$ Faculty of Nuclear Sciences and Physical Engineering, Czech Technical University in Prague, Prague, Czech Republic\\
$^{35}$ Faculty of Physics, Sofia University, Sofia, Bulgaria\\
$^{36}$ Faculty of Science, P.J.~\v{S}af\'{a}rik University, Ko\v{s}ice, Slovak Republic\\
$^{37}$ Faculty of Technology, Environmental and Social Sciences, Bergen, Norway\\
$^{38}$ Frankfurt Institute for Advanced Studies, Johann Wolfgang Goethe-Universit\"{a}t Frankfurt, Frankfurt, Germany\\
$^{39}$ Fudan University, Shanghai, China\\
$^{40}$ Gangneung-Wonju National University, Gangneung, Republic of Korea\\
$^{41}$ Gauhati University, Department of Physics, Guwahati, India\\
$^{42}$ Helmholtz-Institut f\"{u}r Strahlen- und Kernphysik, Rheinische Friedrich-Wilhelms-Universit\"{a}t Bonn, Bonn, Germany\\
$^{43}$ Helsinki Institute of Physics (HIP), Helsinki, Finland\\
$^{44}$ High Energy Physics Group,  Universidad Aut\'{o}noma de Puebla, Puebla, Mexico\\
$^{45}$ Horia Hulubei National Institute of Physics and Nuclear Engineering, Bucharest, Romania\\
$^{46}$ HUN-REN Wigner Research Centre for Physics, Budapest, Hungary\\
$^{47}$ Indian Institute of Technology Bombay (IIT), Mumbai, India\\
$^{48}$ Indian Institute of Technology Indore, Indore, India\\
$^{49}$ INFN, Laboratori Nazionali di Frascati, Frascati, Italy\\
$^{50}$ INFN, Sezione di Bari, Bari, Italy\\
$^{51}$ INFN, Sezione di Bologna, Bologna, Italy\\
$^{52}$ INFN, Sezione di Cagliari, Cagliari, Italy\\
$^{53}$ INFN, Sezione di Catania, Catania, Italy\\
$^{54}$ INFN, Sezione di Padova, Padova, Italy\\
$^{55}$ INFN, Sezione di Pavia, Pavia, Italy\\
$^{56}$ INFN, Sezione di Torino, Turin, Italy\\
$^{57}$ INFN, Sezione di Trieste, Trieste, Italy\\
$^{58}$ Inha University, Incheon, Republic of Korea\\
$^{59}$ Institute for Gravitational and Subatomic Physics (GRASP), Utrecht University/Nikhef, Utrecht, Netherlands\\
$^{60}$ Institute of Experimental Physics, Slovak Academy of Sciences, Ko\v{s}ice, Slovak Republic\\
$^{61}$ Institute of Physics, Homi Bhabha National Institute, Bhubaneswar, India\\
$^{62}$ Institute of Physics of the Czech Academy of Sciences, Prague, Czech Republic\\
$^{63}$ Institute of Space Science (ISS), Bucharest, Romania\\
$^{64}$ Institut f\"{u}r Kernphysik, Johann Wolfgang Goethe-Universit\"{a}t Frankfurt, Frankfurt, Germany\\
$^{65}$ Instituto de Ciencias Nucleares, Universidad Nacional Aut\'{o}noma de M\'{e}xico, Mexico City, Mexico\\
$^{66}$ Instituto de F\'{i}sica, Universidade Federal do Rio Grande do Sul (UFRGS), Porto Alegre, Brazil\\
$^{67}$ Instituto de F\'{\i}sica, Universidad Nacional Aut\'{o}noma de M\'{e}xico, Mexico City, Mexico\\
$^{68}$ iThemba LABS, National Research Foundation, Somerset West, South Africa\\
$^{69}$ Jeonbuk National University, Jeonju, Republic of Korea\\
$^{70}$ Johann-Wolfgang-Goethe Universit\"{a}t Frankfurt Institut f\"{u}r Informatik, Fachbereich Informatik und Mathematik, Frankfurt, Germany\\
$^{71}$ Korea Institute of Science and Technology Information, Daejeon, Republic of Korea\\
$^{72}$ Laboratoire de Physique Subatomique et de Cosmologie, Universit\'{e} Grenoble-Alpes, CNRS-IN2P3, Grenoble, France\\
$^{73}$ Lawrence Berkeley National Laboratory, Berkeley, California, United States\\
$^{74}$ Lund University Department of Physics, Division of Particle Physics, Lund, Sweden\\
$^{75}$ Nagasaki Institute of Applied Science, Nagasaki, Japan\\
$^{76}$ Nara Women{'}s University (NWU), Nara, Japan\\
$^{77}$ National and Kapodistrian University of Athens, School of Science, Department of Physics , Athens, Greece\\
$^{78}$ National Centre for Nuclear Research, Warsaw, Poland\\
$^{79}$ National Institute of Science Education and Research, Homi Bhabha National Institute, Jatni, India\\
$^{80}$ National Nuclear Research Center, Baku, Azerbaijan\\
$^{81}$ National Research and Innovation Agency - BRIN, Jakarta, Indonesia\\
$^{82}$ Niels Bohr Institute, University of Copenhagen, Copenhagen, Denmark\\
$^{83}$ Nikhef, National institute for subatomic physics, Amsterdam, Netherlands\\
$^{84}$ Nuclear Physics Group, STFC Daresbury Laboratory, Daresbury, United Kingdom\\
$^{85}$ Nuclear Physics Institute of the Czech Academy of Sciences, Husinec-\v{R}e\v{z}, Czech Republic\\
$^{86}$ Oak Ridge National Laboratory, Oak Ridge, Tennessee, United States\\
$^{87}$ Ohio State University, Columbus, Ohio, United States\\
$^{88}$ Physics department, Faculty of science, University of Zagreb, Zagreb, Croatia\\
$^{89}$ Physics Department, Panjab University, Chandigarh, India\\
$^{90}$ Physics Department, University of Jammu, Jammu, India\\
$^{91}$ Physics Program and International Institute for Sustainability with Knotted Chiral Meta Matter (WPI-SKCM$^{2}$), Hiroshima University, Hiroshima, Japan\\
$^{92}$ Physikalisches Institut, Eberhard-Karls-Universit\"{a}t T\"{u}bingen, T\"{u}bingen, Germany\\
$^{93}$ Physikalisches Institut, Ruprecht-Karls-Universit\"{a}t Heidelberg, Heidelberg, Germany\\
$^{94}$ Physik Department, Technische Universit\"{a}t M\"{u}nchen, Munich, Germany\\
$^{95}$ Politecnico di Bari and Sezione INFN, Bari, Italy\\
$^{96}$ Research Division and ExtreMe Matter Institute EMMI, GSI Helmholtzzentrum f\"ur Schwerionenforschung GmbH, Darmstadt, Germany\\
$^{97}$ Saga University, Saga, Japan\\
$^{98}$ Saha Institute of Nuclear Physics, Homi Bhabha National Institute, Kolkata, India\\
$^{99}$ School of Physics and Astronomy, University of Birmingham, Birmingham, United Kingdom\\
$^{100}$ Secci\'{o}n F\'{\i}sica, Departamento de Ciencias, Pontificia Universidad Cat\'{o}lica del Per\'{u}, Lima, Peru\\
$^{101}$ Stefan Meyer Institut f\"{u}r Subatomare Physik (SMI), Vienna, Austria\\
$^{102}$ SUBATECH, IMT Atlantique, Nantes Universit\'{e}, CNRS-IN2P3, Nantes, France\\
$^{103}$ Sungkyunkwan University, Suwon City, Republic of Korea\\
$^{104}$ Suranaree University of Technology, Nakhon Ratchasima, Thailand\\
$^{105}$ Technical University of Ko\v{s}ice, Ko\v{s}ice, Slovak Republic\\
$^{106}$ The Henryk Niewodniczanski Institute of Nuclear Physics, Polish Academy of Sciences, Cracow, Poland\\
$^{107}$ The University of Texas at Austin, Austin, Texas, United States\\
$^{108}$ Universidad Aut\'{o}noma de Sinaloa, Culiac\'{a}n, Mexico\\
$^{109}$ Universidade de S\~{a}o Paulo (USP), S\~{a}o Paulo, Brazil\\
$^{110}$ Universidade Estadual de Campinas (UNICAMP), Campinas, Brazil\\
$^{111}$ Universidade Federal do ABC, Santo Andre, Brazil\\
$^{112}$ Universitatea Nationala de Stiinta si Tehnologie Politehnica Bucuresti, Bucharest, Romania\\
$^{113}$ University of Cape Town, Cape Town, South Africa\\
$^{114}$ University of Derby, Derby, United Kingdom\\
$^{115}$ University of Houston, Houston, Texas, United States\\
$^{116}$ University of Jyv\"{a}skyl\"{a}, Jyv\"{a}skyl\"{a}, Finland\\
$^{117}$ University of Kansas, Lawrence, Kansas, United States\\
$^{118}$ University of Liverpool, Liverpool, United Kingdom\\
$^{119}$ University of Science and Technology of China, Hefei, China\\
$^{120}$ University of South-Eastern Norway, Kongsberg, Norway\\
$^{121}$ University of Tennessee, Knoxville, Tennessee, United States\\
$^{122}$ University of the Witwatersrand, Johannesburg, South Africa\\
$^{123}$ University of Tokyo, Tokyo, Japan\\
$^{124}$ University of Tsukuba, Tsukuba, Japan\\
$^{125}$ Universit\"{a}t M\"{u}nster, Institut f\"{u}r Kernphysik, M\"{u}nster, Germany\\
$^{126}$ Universit\'{e} Clermont Auvergne, CNRS/IN2P3, LPC, Clermont-Ferrand, France\\
$^{127}$ Universit\'{e} de Lyon, CNRS/IN2P3, Institut de Physique des 2 Infinis de Lyon, Lyon, France\\
$^{128}$ Universit\'{e} de Strasbourg, CNRS, IPHC UMR 7178, F-67000 Strasbourg, France, Strasbourg, France\\
$^{129}$ Universit\'{e} Paris-Saclay, Centre d'Etudes de Saclay (CEA), IRFU, D\'{e}partment de Physique Nucl\'{e}aire (DPhN), Saclay, France\\
$^{130}$ Universit\'{e}  Paris-Saclay, CNRS/IN2P3, IJCLab, Orsay, France\\
$^{131}$ Universit\`{a} degli Studi di Foggia, Foggia, Italy\\
$^{132}$ Universit\`{a} del Piemonte Orientale, Vercelli, Italy\\
$^{133}$ Universit\`{a} di Brescia, Brescia, Italy\\
$^{134}$ Variable Energy Cyclotron Centre, Homi Bhabha National Institute, Kolkata, India\\
$^{135}$ Warsaw University of Technology, Warsaw, Poland\\
$^{136}$ Wayne State University, Detroit, Michigan, United States\\
$^{137}$ Yale University, New Haven, Connecticut, United States\\
$^{138}$ Yildiz Technical University, Istanbul, Turkey\\
$^{139}$ Yonsei University, Seoul, Republic of Korea\\
$^{140}$ Affiliated with an institute covered by a cooperation agreement with CERN\\
$^{141}$ Affiliated with an international laboratory covered by a cooperation agreement with CERN.\\

\end{flushleft} 
  
\end{document}